\documentclass[aps,amsmath,amssymb,prd,showpacs,floatfix,preprint,superscriptaddress,nofootinbib,12pt]{article}
\usepackage{jheppub}
\usepackage{bm}
\usepackage{ulem}

\usepackage{axodraw4j}
\usepackage{pstricks}
\usepackage{color}

\allowdisplaybreaks[3]

\begin{document}

\title{Conformal correlators in the critical $O(N)$ vector model}

\author[]{Noam Chai,}
\author[]{Mikhail Goykhman,}
\author[]{and Ritam Sinha}
\affiliation[]{The Racah Institute of Physics, The Hebrew University of Jerusalem, \\ Jerusalem 91904, Israel}
\emailAdd{noam.chai@mail.huji.ac.il}
\emailAdd{michael.goykhman@mail.huji.ac.il}
\emailAdd{ritam.sinha@mail.huji.ac.il}

\abstract{
We calculate a set of conformal correlators in the critical $O(N)$ vector model in $2<d<6$ dimensions. We focus on the correlators involving the Hubbard-Stratonovich field $s$, and its composite form $s^2$. In the process, we report a number of new calculations of
diagrams involving the composite $s^2$ operator. Through the calculation of the $\langle s^2s^2s\rangle$ three-point function, we shed new light on a conjectured $s\rightarrow -s$ symmetry in the $s$ sector of the critical $O(N)$ vector model in $d=3$.
}

\maketitle

\section{Introduction}
\label{sec:introduction}

The critical $O(N)$ vector model with quartic interaction and the critical $U(n)$ Gross-Neveu model
are some of the most well-studied interacting conformal field theories.
These CFTs are free in even space-time dimensions, which allows
one to study them perturbatively via the $\epsilon$-expansion in the Wilson-Fisher regime \cite{Wilson:1971dc,Gross:1974jv,Brezin:1976qa,ZinnJustin:1991yn,Fei:2014yja,Fei:2014xta}. 
In three dimensions, these models are strongly coupled and are not accessible to the perturbative treatment, but
prove to be of great interest from the standpoint of understanding the behavior of quantum systems at criticality.
Other methods, such as the $1/N$ expansion \cite{Vasiliev:1981yc,Vasiliev:1981dg,Gracey:1990wi,Gracey:1992cp,Derkachov:1993uw,Vasiliev:1992wr,Vasiliev:1993pi,Gracey:1993kb,Gracey:1993kc,Manashov:2016uam,Manashov:2017rrx} and the conformal bootstrap \cite{Parisi:1972zm,Polyakov:1974gs,Ferrara:1973yt,ElShowk:2012ht,Simmons-Duffin:2016gjk} are frequently used to study such CFTs, and have recently been undergoing an active development.

The interest in these critical models in $d=3$ dimensions is greatly amplified by their
relevance in the context of the $3d/4d$ holographic duality, where they are
described by the dual higher-spin theories in the AdS bulk \cite{Klebanov:2002ja}.\footnote{See also
\cite{Petkou:2003zz,Sezgin:2003pt,Giombi:2009wh} and references therein for some of the earlier works.}
When coupled to the Chern-Simons field, the fundamental scalar and fermionic matter exhibit
interesting Bose/Fermi dualities, and its holographic dual, in turn, has been conjectured to be an interpolation
between type A and type B Vasiliev higher-spin theories \cite{Giombi:2011kc,Aharony:2011jz,Aharony:2012nh}.

In the deep UV regime, the Gross-Neveu model in $2\leq d\leq 4$ dimensions
reaches a fixed point. This fixed point can be studied perturbatively 
in the vicinity of $d=2$ dimensions (where the model is asymptotically free \cite{Gross:1974jv}),
as well as in the vicinity of $d=4$ dimensions (where the model is critically
equivalent to the Gross-Neveu-Yukawa model \cite{ZinnJustin:1991yn}).
In general $d$, this interacting CFT describes
dynamics of the fermions $\psi^i$ and the Hubbard-Stratonovich scalar field $\sigma$,
and its action is given by
\begin{equation}
\label{GN action}
S_{\textrm{G.N}} = \int d^d x\,\left(\bar\psi^i\gamma^\mu\partial_\mu\psi^i +\frac{1}{\sqrt{N}}\,\sigma
\bar\psi^i\psi^i\right)\,.
\end{equation}
The Gross-Neveu model action (\ref{GN action}) is invariant w.r.t. the discrete $\mathbb{Z}_2$ symmetry \cite{Gross:1974jv,ZinnJustin:1991yn,Moshe:2003xn}
\begin{align}
&(x^1,\dots, x^{a-1},x^a,x^{a+1},\dots, x^d) \rightarrow
(x^1,\dots, x^{a-1}, - x^a,x^{a+1},\dots, x^d)\,,\notag\\
&\sigma\rightarrow -\sigma\,,\quad \psi\rightarrow \gamma_a \psi\,,\quad \bar\psi\rightarrow -\bar\psi\gamma_a\,,
\label{Z2 in GN}
\end{align}
for any given $a=1,\dots ,d$. In particular, (\ref{Z2 in GN}) implies $\bar\psi\psi\rightarrow -\bar\psi\psi$,
and as a result the interaction term $\sigma \bar\psi^i\psi ^i$ in the action (\ref{GN action}) is left invariant.

Furthermore, the symmetry (\ref{Z2 in GN}) uniquely fixes the structure of some
of the correlation functions in the critical GN model. For instance, while conformal
symmetry allows two possible structures of the three-point correlation functions
involving one scalar and two fermionic fields \cite{Weinberg:2010fx,Goykhman:2018ihr},
the $\mathbb{Z}_2$ symmetry (\ref{Z2 in GN}) can further select which one of these two
structures is allowed, as can be seen on the following example \cite{Goykhman:2020ffn}:
\begin{align}
\langle \bar\psi(x_1)\psi(x_2)\sigma(x_3)^{2k+1}\rangle &\sim\frac{\gamma_\mu\, x_{13}^\mu\,
\gamma_\nu \, x_{32}^\nu}{|x_{12}|^{d-2-2k}(|x_{13}||x_{23}|)^{2k+2}}\,,\label{psi psi s2kp1}\\
\langle \bar\psi(x_1)\psi(x_2)\sigma(x_3)^{2(k+1)}\rangle &\sim\frac{\gamma_\mu\, x_{12}^\mu}
{|x_{12}|^{d-2(k+1)}(|x_{13}||x_{23}|)^{2(k+1)}}\,,\label{psi psi s2kp2}
\end{align}
where $k=0,1,2,\dots$. While (\ref{psi psi s2kp1}), (\ref{psi psi s2kp2})
are leading order in $1/N$, the same structure will hold exactly to all orders in the $1/N$ expansion.

One can also study manifestations of the symmetry (\ref{Z2 in GN}) in the singlet
sector of the critical GN model. This is particularly relevant from the standpoint
of the holographic duality, which provides a prescription for evaluation of the 
correlation function of the $U(n)$ singlets via the dual description of the gauge degrees of freedom in the AdS bulk.
An immediate consequence of the symmetry (\ref{Z2 in GN})
is that the correlation functions involving an odd number of the Hubbard-Stratonovich
fields $\sigma$ vanish. The simplest example of this statement is given by the triviality of 
the three-point function
\begin{equation}
\label{vanishing 3pt function in GN model}
\langle \sigma(x_1) \sigma(x_2) \sigma(x_3)\rangle = 0\,.
\end{equation}
The relation (\ref{vanishing 3pt function in GN model}) can be explicitly verified 
by evaluating the corresponding Feynman diagrams \cite{Manashov:2016uam,Goykhman:2020ffn}.\\

In this paper, we study the critical $O(N)$ vector model with the action 
\begin{equation}
\label{starting action intro}
S = \int d^dx \,\left(\frac{1}{2}\,\partial_\mu\phi^i\,\partial^\mu\phi_i + \frac{1}{\sqrt{N}}\,s\,\phi^i\phi^i\right)\,,
\end{equation}
describing dynamics of the fundamental fields $\phi^i$ and the Hubbard-Stratonovich field $s$.
At first glance, the model (\ref{starting action intro}) does not seem to possess any such remarkable discrete $\mathbb{Z}_2$ symmetry due to the purely bosonic nature of the fields $\phi_i$. However, a conjecture was put forth in \cite{Petkou:1994ad} regarding the existence of such a symmetry in the $O(N)$ vector models in $d=3$, based on conformal bootstrap calculation of the 
three-point function $\langle sss\rangle$ at the leading order in $1/N$ expansion.
It is our aim to shed further light on the fate of this conjecture at both the leading and next-to leading order in $1/N$
in the model (\ref{starting action intro}).
Inspired by the GN model, we will search for manifestations of this symmetry in certain conformal correlators.
We will be mostly interested in the $d=3$ case, but majority of our calculations will be carried out in general $d$.

The CFT action (\ref{starting action intro}) describes the critical behavior of the $O(N)$ vector model with quartic interaction
at its IR fixed point, and the non-linear sigma model at its UV fixed point, for $2<d<4$ \cite{ZinnJustin:1991yn}.
When $4<d<6$ the model (\ref{starting action}) describes a UV fixed point of the $O(N)$
vector model with quartic interaction, and is conjectured to describe the IR fixed point of
certain vector model with cubic coupling. The latter statement is supported by a perturbative calculation
in $d=6-\epsilon$ dimensions up to quartic order \cite{Fei:2014yja,Fei:2014xta,Gracey:2015tta}.

The counterpart of the transformation (\ref{Z2 in GN}) in the $O(N)$ vector model (\ref{starting action intro})
would only act on the Hubbard-Stratonovich field $s$ \cite{Petkou:1994ad},
\begin{equation}
\label{Z2 in ON}
s\rightarrow - s\,.
\end{equation}
While the transformation (\ref{Z2 in ON}) is clearly not a symmetry of the action (\ref{starting action intro}),
it has been suggested in \cite{Petkou:1994ad} that such a symmetry might emerge
in the $d=3$ dimensional quantum theory among the correlation functions involving only the $s$ fields. Interestingly, \cite{Petkou:1994ad} pointed out that the three-point function
vanishes at the leading order in the $1/N$ expansion,
\begin{equation}
\label{vanishing leading 3pt function in ON model}
\langle s(x_1) s(x_2) s(x_3)\rangle\Bigg|_{d=3} = 0+ {\cal O}\left(\frac{1}{N^{3/2}}\right)\,,
\end{equation}
and suggested to explain it by conjecturing the symmetry (\ref{Z2 in ON}).
Notice that (\ref{vanishing leading 3pt function in ON model}) is not valid when $d\neq 3$,
in stark contrast with the GN case, where the $\mathbb{Z}_2$ symmetry holds for any $d$.
In \cite{Goykhman:2019kcj} the $\langle sss\rangle$ correlation
function was calculated at the next-to-leading order in the $1/N$ expansion. Remarkably,
\cite{Goykhman:2019kcj} demonstrated that the sub-leading correction to the three-point function $\langle sss\rangle$ also
vanishes in $d=3$,
\begin{equation}
\label{vanishing next to leading 3pt function in ON model}
\langle s(x_1) s(x_2) s(x_3)\rangle\Bigg|_{d=3} = 0+ {\cal O}\left(\frac{1}{N^{5/2}}\right)\,.
\end{equation}
This further raises the question of whether the symmetry \eqref{Z2 in ON} is valid also upto the first sub-leading order in $1/N$.

The proposed symmetry, however, is fundamentally different than its counterpart (\ref{Z2 in GN})  in the GN model for a number of reasons. Primary among them is that the symmetry is present only in $d=3$. However, more importantly, the conjectured symmetry transformation (\ref{Z2 in ON}) is not respected by any correlation functions involving the fundamental scalar $\phi_i$. As a simple example, one can notice that
the correlation
function $\langle \phi\phi s\rangle$ is non-vanishing in $d=3$ \cite{Petkou:1994ad}, although
it is odd w.r.t. the transformation (\ref{Z2 in ON}). In fact, it was originally suggested in
\cite{Petkou:1994ad} that the symmetry (\ref{Z2 in ON}), if established, would have to be confined to the sub-sector of the theory, involving only the $s$  fields. This is very unlike the GN model where correlations involving the fundamental $\psi_i$ field also respect the discrete symmetry (\ref{Z2 in GN}), as we reviewed above on the example of the
correlation functions (\ref{psi psi s2kp1}), (\ref{psi psi s2kp2}).

One might ask if such a limited scope of applicability of the conjectured symmetry transformation 
(\ref{Z2 in ON}) can be explained by the holographic dual of the three-dimensional vector model, that
describes only the singlet sector of the theory. To test such an assumption, one can 
consider the three-point function $\langle \phi^2 \phi^2 s\rangle$. At the leading order in $1/N$,
this correlation function is given by the diagram\footnote{Our notations and Feynman rules
will be explained in section~\ref{sec:setup}.}
\begin{center}
  \begin{picture}(279,146) (24,0)
    \SetWidth{1.0}
    \SetColor{Black}
    \Line[](98,135)(98,72)
    \Line[](98,72)(42,16)
    \Line[](98,72)(154,16)
    \Line[](42,16)(154,16)
    \Vertex(98,72){4}
    \Vertex(98,135){2}
    \Vertex(42,16){2}
    \Vertex(154,16){2}
    \Text(21,-3)[lb]{\scalebox{0.8}{$\phi(x_1)^2$}}
    \Text(160,-3)[lb]{\scalebox{0.8}{$\phi(x_2)^2$}}
    \Text(104,135)[lb]{\scalebox{0.8}{$s(x_3)$}}
    \Text(105,100)[lb]{\scalebox{0.8}{$2\Delta_s$}}
    \Text(52,44)[lb]{\scalebox{0.8}{$2\Delta_\phi$}}
    \Text(132,44)[lb]{\scalebox{0.8}{$2\Delta_\phi$}}
    \Text(91,2)[lb]{\scalebox{0.8}{$2\Delta_\phi$}}
    \Text(190,64)[lb]{\scalebox{1}{$=-\frac{1}{8\pi^5\sqrt{N}}\,\frac{1}{(|x_{13}||x_{23}|)^2}\qquad \textrm{(in d=3)}$}}
  \end{picture}
\end{center}
which we evaluated in $3d$, obtaining a non-zero result. This quick calculation further
excises the scope of the proposed symmetry transformation (\ref{Z2 in ON}), removing the $O(N)$ singlet
composite operators involving the field $\phi_i$.

We then intend to study the transformation of the three-point correlation functions
involving the Hubbard-Stratonovich field $s$ only.
To test the proposed symmetry transformation (\ref{Z2 in ON}) one needs to study correlation
functions involving an odd number of the fields $s$. Above we have discussed that the three-point function
of the $s$ field vanishes up to the next-to-leading order in $1/N$ expansion. A natural next step is to study three-point
functions involving five of the fields $s$, which requires one to deal with the composite operators $s^2$ or $s^3$. 

Specifically, the objective of this paper
is to calculate the three-point correlation function $\langle s^2 s^2 s\rangle$.
While we establish that this correlator vanishes at the leading order in the $1/N$
expansion, our main result is that its next-to-leading order correction is in fact non-zero,
and therefore does not respect the conjectured symmetry (\ref{Z2 in ON}).

In the process we obtain a number of new results, which can be used for other calculations in the $O(N)$
vector model. Some of the new expressions obtained in this paper involve the $s^2 s s$
conformal triangle at the next-to-leading order in $1/N$ in general $d$, the $\langle s^2 s s\rangle$
three-point function in $d=3$, and several self-energy diagrams, which we believe have not been reported
in the literature before. The rest of this paper is organized as follows.
In section~\ref{sec:setup} we set up our notations and review known results in the critical $O(N)$
vector model that will be relevant for the purposes of this paper. We also 
reformulate the result of \cite{Goykhman:2019kcj} for the $\langle sss\rangle$ correlation function in terms
of the $sss$ conformal triangle, representing the cubic effective vertex at the next-to-leading order in $1/N$.
In section~\ref{sec:s2ss} we derive the $s^2 s s$ conformal triangle in general $d$,
and calculate the $\langle s^2 ss\rangle$ three-point function in $d=3$ at the next-to-leading order in $1/N$.
In section~\ref{s2s2s} we derive the $\langle s^2s^2s\rangle$ correlation function in $d=3$. We demonstrate
that while the leading-order $3d$ correlation function vanishes, its $1/N$ correction is non-trivial.
Motivated by this result, in section~\ref{sec:leading order} we then explore whether the 
conjectured symmetry (\ref{Z2 in ON}) is an artifact of the large-$N$ limit.
We discuss our results in section~\ref{sec:discussion}.

\section{Set-up}
\label{sec:setup}

In this paper we will be studying the critical $O(N)$-invariant vector model with the action
\begin{equation}
\label{starting action}
S = \int d^dx \,\left(\frac{1}{2}\,(\partial\phi)^2 + \frac{1}{
\sqrt{N}}\,s\,\phi^2\right) + S_{\textrm{c.t.}}\,,
\end{equation}
describing dynamics of the multiplet $\phi^i$, $i = 1,\dots, N$
of the real-valued scalar fields, and the Hubbard-Stratonovich field $s$.\footnote{Here and in
what follows we skip keeping track of the $O(N)$ indices where it does not cause a confusion.}
In the action (\ref{starting action}) we have also incorporated the counterterm induced by
wave-function renormalization of the fields $\phi$, $s$, see \cite{Goykhman:2019kcj} for a recent review.

The $\phi$ field propagator is given by
\begin{equation}
\label{phi propagator}
\langle \phi(x) \phi(0) \rangle = \frac{C_{\phi}\,(1+A_\phi)\,\mu^{-2\gamma_\phi}}{|x|^{2(\Delta_\phi
+\gamma_\phi)}}\,,
\end{equation}
where $\mu$ is an arbitrary RG scale,
$\Delta_\phi$ is the free scaling dimension, and $\gamma_\phi$ is the anomalous dimension, given by \cite{Vasiliev:1981yc,Vasiliev:1981dg,Petkou:1994ad}
\begin{align}
\label{full phi dim}
\Delta_\phi &= \frac{d}{2} - 1 \,,\\
\label{phi anomalous dim}
\gamma_\phi &= \frac{1}{N}\,\frac{2^d \sin \left(\frac{\pi  d}{2}\right) \Gamma \left(\frac{d-1}{2}\right)}{\pi ^{3/2} (d-2) d \Gamma \left(\frac{d}{2}-2\right)}+{\cal O}\left(\frac{1}{N^2}\right)\,,
\end{align}
leading in $1/N$ amplitude is given by
\begin{equation}
C_\phi = \frac{\Gamma\left(\frac{d}{2}-1\right)}{4\pi^\frac{d}{2}}\,,
\end{equation}
and sub-leading correction to the amplitude is \cite{Derkachov:1997ch}
\begin{equation}
\label{Aphi}
A_\phi = \left(\frac{d}{2-d} - \frac{2}{d}\right)\,\gamma_\phi +{\cal O}\left(\frac{1}{N^2}\right)\,.
\end{equation}
The Feynman rule corresponding to the propagator (\ref{phi propagator}) is
\begin{center}
  \begin{picture}(147,20) (23,0)
    \SetWidth{1.0}
    \SetColor{Black}
    \Line[](31,5)(112,5)
    \Vertex(31,5){2}
    \Vertex(112,5){2}
    \Text(65,10)[lb]{\scalebox{0.8}{$2\Delta_\phi$}}
    \Text(145,-5)[lb]{\scalebox{1}{$=\frac{C_\phi}{|x|^{2\Delta_\phi}}$}}
    \Text(20,4)[lb]{\scalebox{0.8}{$0$}}
    \Text(122,4)[lb]{\scalebox{0.8}{$x$}}
  \end{picture}
\end{center}

In a general conformal graph we will also use the following Feynman rule
for internal lines with unit amplitudes
\begin{center}
  \begin{picture}(147,20) (23,0)
    \SetWidth{1.0}
    \SetColor{Black}
    \Line[](31,5)(112,5)
    \Vertex(31,5){2}
    \Vertex(112,5){2}
    \Text(65,10)[lb]{\scalebox{0.8}{$2a$}}
    \Text(145,-5)[lb]{\scalebox{1}{$=\frac{1}{|x|^{2a}}$}}
    \Text(20,4)[lb]{\scalebox{0.8}{$0$}}
    \Text(122,4)[lb]{\scalebox{0.8}{$x$}}
  \end{picture}
\end{center}

Since the action (\ref{starting action}) is quadratic in $\phi$, the corresponding
path integral is Gaussian, and can be performed explicitly, resulting in the effective action
for $s$ formally written as
\begin{equation}
S_{\textrm{eff}} = \frac{N}{2}\,\int d^dx\, \textrm{Tr}\log\left(\partial^2 - \frac{2}{\sqrt{N}}\,s\right)\,.
\end{equation}
Expanding the logarithm, we obtain
\begin{equation}
\label{starting effective action for s}
S_{\textrm{eff}} = -C_\phi^2\int d^dx_{1,2}\,\frac{s(x_1)s(x_2)}{|x_{12}|^{2(d-2)}}
+\frac{4C_\phi^3}{3\sqrt{N}}\,\int d^dx_{1,2,3}
\,\frac{s(x_1)s(x_2)s(x_3)}{(|x_{12}||x_{13}||x_{23}|)^{d-2}}+\cdots\,,
\end{equation}
where ellipsis stand for vertices of higher order in $s$.

From the quadratic term in the action (\ref{starting effective action for s}) we obtain
the propagator for the Hubbard-Stratonovich field $s$,
\begin{equation}
\label{free s propagator}
\langle s(x) s(0) \rangle = \frac{C_{s}}{|x|^{2\Delta_s}}\,,
\end{equation}
where 
\begin{align}
\Delta_s &=2\,,\\
C_s &= \frac{2^d\,\Gamma\left(\frac{d-1}{2}\right)\sin\left(\frac{\pi d}{2}\right)}{\pi^\frac{3}{2}\Gamma\left(
\frac{d}{2}-2\right)}\,.
\end{align}
The corresponding Feynman rule is
\begin{center}
  \begin{picture}(147,20) (23,0)
    \SetWidth{1.0}
    \SetColor{Black}
    \Line[](31,5)(112,5)
    \Vertex(31,5){2}
    \Vertex(112,5){2}
    \Text(65,10)[lb]{\scalebox{0.8}{$2\Delta_s$}}
    \Text(145,-5)[lb]{\scalebox{1}{$=\frac{C_s}{|x|^{2\Delta_s}}$}}
    \Text(20,4)[lb]{\scalebox{0.8}{$0$}}
    \Text(122,4)[lb]{\scalebox{0.8}{$x$}}
  \end{picture}
\end{center}

The loop corrections to the Hubbard-Stratonovich propagator result in 
\begin{equation}
\label{s propagator}
\langle s(x) s(0) \rangle = \frac{C_{s}\,(1+A_s)\,\mu^{-2\gamma_s}}{|x|^{2(\Delta_s+\gamma_s)}}\,,
\end{equation}
where \cite{Vasiliev:1981yc,Vasiliev:1981dg,Petkou:1994ad,Derkachov:1997ch}
\begin{align}
\label{s anomalous dim}
\gamma_s &= 4 \left(d+\frac{6}{d-4}+1\right)\,\gamma_\phi+{\cal O}\left(\frac{1}{N^2}\right)\,,\\
\label{As}
A_s &{=} 2\gamma_\phi \left(\frac{d(d{-}3){+}4}{4{-}d}\,\left(H_{d{-}3}{+}\pi\cot\left(\frac{\pi d}{2}\right)\right) 
{+}\frac{8}{(d{-}4)^2}{+}\frac{2}{d{-}2}{+}\frac{2}{d}{-}2d{-}1\right){+}{\cal O}\left(\frac{1}{N^2}\right)\,.
\end{align}
We will also use the Feynman rules corresponding to the dressed propagator
\begin{center}
  \begin{picture}(147,20) (23,0)
    \SetWidth{1.0}
    \SetColor{Black}
    \Line[](31,5)(112,5)
    \GOval(73,5)(7,7)(0){0.882}
    \Vertex(31,5){2}
    \Vertex(112,5){2}
    \Text(72,15)[lb]{\scalebox{0.8}{$s$}}
    \Text(145,-5)[lb]{\scalebox{1}{$=\frac{C_s\,(1+A_s)\,\mu^{-2\gamma_s}}{|x|^{2(\Delta_s+\gamma_s)}}$}}
    \Text(20,4)[lb]{\scalebox{0.8}{$0$}}
    \Text(122,4)[lb]{\scalebox{0.8}{$x$}}
  \end{picture}
\end{center}
\begin{center}
  \begin{picture}(147,20) (23,0)
    \SetWidth{1.0}
    \SetColor{Black}
    \Line[](31,5)(112,5)
    \Vertex(73,5){8}
    \Vertex(31,5){2}
    \Vertex(112,5){2}
    \Text(72,15)[lb]{\scalebox{0.8}{$s$}}
    \Text(145,-5)[lb]{\scalebox{1}{$=\frac{C_s\,\mu^{-2\gamma_s}}{|x|^{2(\Delta_s+\gamma_s)}}$}}
    \Text(20,4)[lb]{\scalebox{0.8}{$0$}}
    \Text(122,4)[lb]{\scalebox{0.8}{$x$}}
  \end{picture}
\end{center}

Higher order terms in the
action (\ref{starting effective action for s}) can be represented by conformal graphs with internal $\phi$ lines
and the interaction vertex
\begin{center}
  \begin{picture}(169,58) (43,22)
    \SetWidth{1.0}
    \SetColor{Black}
    \Line[](42,85)(42,42)
    \Line[](42,42)(73,19)
    \Line[](42,42)(13,18)
    \Vertex(42,42){4}
    \Text(47,71)[lb]{\scalebox{0.8}{$s$}}
    \Text(10,25)[lb]{\scalebox{0.8}{$\phi$}}
    \Text(70,25)[lb]{\scalebox{0.8}{$\phi$}}
    \Text(127,36)[lb]{\scalebox{1}{$=-\frac{2}{\sqrt{N}}$}}
  \end{picture}
\end{center}
In  particular, second term in the r.h.s. of (\ref{starting effective action for s})
gives the leading ${\cal O}(1/N^{1/2})$ order $sss$ vertex. Sub-leading corrections
to this vertex can be written down in terms of the corresponding Polyakov's \cite{Polyakov:1974gs} conformal triangle
\begin{equation}
\label{sss vertex}
S_{\textrm{eff}} \supset \frac{Z_{sss}}{\sqrt{N}}\,\int d^dx_{1,2,3}
\,\frac{s(x_1)s(x_2)s(x_3)}{(|x_{12}||x_{13}||x_{23}|)^{d-2-\gamma_s}}\,\mu^{3\gamma_s}\,.
\end{equation}
Here the amplitude of the conformal triangle $Z_{sss}$ and the anomalous
dimension $\gamma_s$ are assumed to be expanded to the desired power in $1/N$,
\begin{equation}
\label{Zsss expansion}
Z_{sss} = Z_{sss}^{(0)} \,(1 + \delta Z_{sss})\,.
\end{equation}
The Feynman rule corresponding to the vertex/conformal triangle (\ref{sss vertex}) is given by
\begin{center}
  \begin{picture}(236,72) (12,17)
    \SetWidth{1.0}
    \SetColor{Black}
    \Line[](144,32)(191,32)
    \Line[](167,69)(145,33)
    \Line[](167,69)(191,33)
    \Line[](143,31)(125,20)
    \Line[](191,32)(209,20)
    \Line[](167,71)(167,88)
    \Line[](42,85)(42,42)
    \Line[](42,42)(73,18)
    \Line[](42,42)(13,18)
    \GOval(42,42)(7,7)(0){0.882}
    \Vertex(167,68){4}
    \Vertex(145,32){4}
    \Vertex(191,32){4}
    \Text(85,45)[lb]{\scalebox{1}{$=-\frac{6Z_{sss}}{\sqrt{N}}\times$}}
    \Text(186,50)[lb]{\scalebox{0.8}{$\alpha$}}
    \Text(143,50)[lb]{\scalebox{0.8}{$\alpha$}}
    \Text(166,25)[lb]{\scalebox{0.8}{$\alpha$}}
    \Text(47,71)[lb]{\scalebox{0.8}{$s$}}
    \Text(10,25)[lb]{\scalebox{0.8}{$s$}}
    \Text(70,25)[lb]{\scalebox{0.8}{$s$}}
    \Text(173,88)[lb]{\scalebox{0.8}{$s$}}
    \Text(119,25)[lb]{\scalebox{0.8}{$s$}}
    \Text(213,25)[lb]{\scalebox{0.8}{$s$}}
  \end{picture}
\end{center}
where we denoted
\begin{equation}
\alpha=d-2-\gamma_s\,.
\end{equation}
We will also use the following notation for the $sss$ conformal triangle with the leading order amplitude only:
\begin{center}
  \begin{picture}(236,72) (12,17)
    \SetWidth{1.0}
    \SetColor{Black}
    \Line[](144,32)(191,32)
    \Line[](167,69)(145,33)
    \Line[](167,69)(191,33)
    \Line[](143,31)(125,20)
    \Line[](191,32)(209,20)
    \Line[](167,71)(167,88)
    \Line[](42,85)(42,42)
    \Line[](42,42)(73,18)
    \Line[](42,42)(13,18)
    \COval(42,42)(7,7)(0){Black}{Black}
    \Vertex(167,68){4}
    \Vertex(145,32){4}
    \Vertex(191,32){4}
    \Text(85,45)[lb]{\scalebox{1}{$=-\frac{6Z_{sss}^{(0)}}{\sqrt{N}}\times$}}
    \Text(186,50)[lb]{\scalebox{0.8}{$\alpha$}}
    \Text(143,50)[lb]{\scalebox{0.8}{$\alpha$}}
    \Text(166,25)[lb]{\scalebox{0.8}{$\alpha$}}
    \Text(47,71)[lb]{\scalebox{0.8}{$s$}}
    \Text(10,25)[lb]{\scalebox{0.8}{$s$}}
    \Text(70,25)[lb]{\scalebox{0.8}{$s$}}
    \Text(173,88)[lb]{\scalebox{0.8}{$s$}}
    \Text(119,25)[lb]{\scalebox{0.8}{$s$}}
    \Text(213,25)[lb]{\scalebox{0.8}{$s$}}
  \end{picture}
\end{center}
Comparing the leading order coefficients of the cubic vertex in (\ref{starting effective action for s})
and (\ref{sss vertex}), we obtain the leading order amplitude $Z_{sss}^{(0)}$ of the conformal
triangle (\ref{Zsss expansion})
\begin{equation}
\label{Zsss0}
Z_{sss}^{(0)} = \frac{4}{3}\, C_\phi^3\,.
\end{equation}
To calculate the sub-leading contribution $\delta Z_{sss}$ to the $sss$
conformal triangle, we use the expression for the three-point function
\begin{equation}
\langle s(x_1)s(x_2)s(x_3)\rangle = \frac{C_{sss}^{(0)}(1+\delta C_{sss})}
{(|x_{12}||x_{13}||x_{23}|)^{\Delta_s+\gamma_s}}\,\mu^{-3\gamma_s}\,,
\end{equation}
where \cite{Petkou:1994ad}\footnote{See App.~\ref{appA} for our conventions.}
\begin{align}
C_{sss}^{(0)}&=N\left(-\frac{2}{\sqrt{N}}\right)^3C_\phi^3 C_s^3U\left(\frac{d}{2}-1,\frac{d}{2}-1,2\right)^2U(1,2,d-3)\\
&=-\frac{1}{\sqrt{N}}\frac{8^{d-1} \sin ^3\left(\frac{\pi  d}{2}\right) \Gamma \left(3-\frac{d}{2}\right) \Gamma \left(\frac{d-1}{2}\right)^3}{\pi ^{9/2} \Gamma (d-3)}\,,
\end{align}
and $\delta C_{sss}$ was found in \cite{Goykhman:2019kcj} at the next-to-leading order in $1/N$\footnote{
In notations of \cite{Goykhman:2019kcj} we have $\delta C_{sss} = W_{s^3}+3A_s/2$. This is because $W_{s^3}$ was defined in \cite{Goykhman:2019kcj} 
as a relative correction to the amplitude of the three-point function of the normalized fields $s$, related
to the non-normalized fields by the transformation $s\rightarrow \sqrt{C_s(1+A_s)} \,s$. Appearance
of the additional term $3A_s/2$ is clear from such a transformation.}
\begin{align}
\delta C_{sss} &= 3W_{\phi\phi s} + f+ W_3+W_4+\frac{3A_s}{2}\,,\\
W_{\phi\phi s} &= \gamma_\phi\,\left(\frac{d(d-3)+4}{4-d}
\Big( H_{d-3}+\pi  \cot \big(\frac{\pi  d}{2}\big)\Big)
+\frac{16}{(d-4)^2}+\frac{6}{d-4}+\frac{2}{d-2}-2 d+3\right)\,,\notag\\
 f&=\gamma_\phi\,\frac{6(d-1)(d-2)}{d-4}\,\left(H_{d-4}-\frac{2}{d-4}+\pi  \cot \big(\frac{\pi  d}{2}\big) \right)\,,\notag\\
W_3 &=\gamma_\phi {2d(d{-}2)(d{-}3)\over (d{-}4)^2} 
\left( 6 \psi ^{(1)}\Big(\frac{d}{2}{-}1\Big){-}\psi ^{(1)}(d{-}3) {-}{\pi^2\over 6}
{-} H_{d-4}\Big(H_{d{-}4}{+}2\pi\cot\big({\pi d\over 2}\big) \Big)
\right) \,,\notag\\
W_4&=\gamma_\phi\,\frac{3 d (d-2)  \left(\pi ^2-6 \psi ^{(1)}\left(\frac{d}{2}-1\right)\right)}{4 (d-4)}\,,
\notag
\end{align}
where $\psi^{(1)}$ is the first derivative of the digamma function, and $H_n$ is the $n$the harmonic number.
On the other hand, the $\langle sss\rangle$ three-point function can be obtained
by attaching three $s$ legs to the $sss$ conformal triangle, and integrating over its vertices:
\begin{center}
  \begin{picture}(236,72) (12,17)
    \SetWidth{1.0}
    \SetColor{Black}
    \Line[](42,85)(42,42)
    \Line[](42,42)(73,18)
    \Line[](42,42)(13,18)
    \GOval(42,42)(7,7)(0){0.882}
    \GOval(42,65)(7,7)(0){0.882}
    \GOval(57,30)(7,7)(0){0.882}
    \GOval(27,30)(7,7)(0){0.882}
    \Vertex(42,85){2}
    \Vertex(13,18){2}
    \Vertex(73,18){2}
    \Text(105,45)[lb]{\scalebox{1.2}{$=\frac{C_{sss}^{(0)}(1+\delta C_{sss})}{(|x_{12}||x_{13}||x_{23}|)^{\Delta_s+\gamma_s}}$}}
    \Text(54,64)[lb]{\scalebox{0.8}{$s$}}
    \Text(10,32)[lb]{\scalebox{0.8}{$s$}}
    \Text(70,32)[lb]{\scalebox{0.8}{$s$}}
    \Text(50,84)[lb]{\scalebox{0.8}{$x_3$}}
    \Text(0,12)[lb]{\scalebox{0.8}{$x_1$}}
    \Text(80,12)[lb]{\scalebox{0.8}{$x_2$}}
  \end{picture}
\end{center}
Then we can express
\begin{align}
\label{deltaZsss in terms of deltaCsss}
\delta Z_{sss} = \delta C_{sss} - 3A_s -R_{sss}\,,
\end{align}
where $R_{sss}$ is obtained by expansion of the factor
\begin{align}
U\left(2+\gamma_s,\frac{d-\gamma_s}{2}-1,\frac{d-\gamma_s}{2}-1\right)^2
&U\left(2+\gamma_s,1+\frac{\gamma_s}{2},d-3-\frac{3\gamma_s}{2}\right)\notag\\
& = u_{sss}^{(0)}\left(1+R_{sss}+{\cal O}\left(
\frac{1}{N^2}\right)\right)
\end{align}
originating from taking the integrals over the vertices of the $sss$ conformal triangle,
\begin{align}
u_{sss}^{(0)}&= \frac{8 \pi ^{\frac{3 d}{2}} \Gamma \left(3-\frac{d}{2}\right)}{(d-4)^4 \Gamma (d-4)}\,,\\
R_{sss} &= \frac{1}{N}\frac{6 \sin \left(\frac{\pi  d}{2}\right) \Gamma (d) \left((d-4) H_{d-4}-2 d+\pi  (d-4) \cot \left(\frac{\pi  d}{2}\right)+10\right)}{\pi  (d-4) \Gamma \left(\frac{d}{2}-1\right) \Gamma \left(\frac{d}{2}+1\right)}\,.
\end{align}
Finally notice that
\begin{equation}
Z_{sss}^{(0)} = -\frac{C_{sss}^{(0)}}{6C_s^3u_{sss}^{(0)}}\,,
\end{equation}
which agrees with (\ref{Zsss0})\,.

\section{$\langle s^2 s s\rangle$}
\label{sec:s2ss}

At the leading order the $\langle s^2 s^2\rangle$ propagator is given by
\begin{center}
  \begin{picture}(368,60) (37,10)
    \SetWidth{1.0}
    \SetColor{Black}
    \CBox(45,35)(50,40){Black}{Black}
    \Arc[clock](287.5,-32.976)(91.978,129.496,50.504)
    \Arc[](287.5,117.5)(98.501,-127.161,-52.839)
    \CBox(169,35)(174,40){Black}{Black}
    \CBox(224,35)(229,40){Black}{Black}
    \CBox(345,35)(350,40){Black}{Black}
    \Line[](50,37)(169,37)
    \Text(37,24)[lb]{\scalebox{0.8}{$0$}}
    \Text(176,24)[lb]{\scalebox{0.8}{$x$}}
    \Text(105,43)[lb]{\scalebox{0.8}{$2\Delta_{s^2}$}}
    \Text(200,36)[lb]{\scalebox{0.8}{$=$}}
    \Text(285,64)[lb]{\scalebox{0.8}{$2\Delta_s$}}
    \Text(285,7)[lb]{\scalebox{0.8}{$2\Delta_s$}}
    \Text(370,26)[lb]{\scalebox{1.2}{$=\frac{C_{s^2}}{|x|^{2\Delta_{s^2}}}$}}
    \Text(223,24)[lb]{\scalebox{0.8}{$0$}}
    \Text(355,24)[lb]{\scalebox{0.8}{$x$}}
  \end{picture}
\end{center}
Here the leading order scaling dimension of the composite operator $s^2$ is given by
\begin{equation}
\Delta_{s^2} = 4\,,
\end{equation}
and the leading order propagator amplitude is
\begin{equation}
C_{s^2} = \frac{1}{2}\,2^2\,C_s^2 = 2C_s^2\,,
\end{equation}
where $1/2$ is the symmetry factor, and each factor of 2 comes from the degeneracy of each of the
two $s^2$ insertions.

The loop corrections to the $s^2$ propagator result in 
\begin{equation}
\label{s propagator}
\langle s(x)^2 s(0)^2 \rangle = \frac{C_{s^2}\,(1+A_{s^2})\,\mu^{-2\gamma_{s^2}}}{|x|^{2(\Delta_{s^2}+\gamma_{s^2})}}\,,
\end{equation}
where  \cite{Lang:1992zw}
\begin{align}
\label{s2 anomalous dim}
\gamma_{s^2} &= -4(d-1)^2\,\gamma_\phi+{\cal O}\left(\frac{1}{N^2}\right)\,.
\end{align}

Due to the conformal symmetry the $\langle s^2 s s\rangle$
three-point function takes the form
\begin{equation}
\label{s2ss general}
\langle s(x_1) s(x_2) s^2(x_3)\rangle = \frac{C_{s^2 s s}^{(0)}(1+\delta C_{s^2 s s})}{|x_{12}|^{2\gamma_s-\gamma_{s^2}}
(|x_{13}||x_{23}|)^{4+\gamma_{s^2}}}\,\mu^{-2\gamma_s-\gamma_{s^2}}\,.
\end{equation}
Here the amplitude of the leading order diagram
\begin{center}
  \begin{picture}(152,55) (36,6)
    \SetWidth{1.0}
    \SetColor{Black}
    \Line[](78,47)(41,11)
    \Line[](79,47)(112,11)
    \Vertex(41,11){2}
    \Vertex(112,11){2}
    \CBox(76,47)(81,52){Black}{Black}
    \Text(44,32)[lb]{\scalebox{0.8}{$2\Delta_s$}}
    \Text(97,32)[lb]{\scalebox{0.8}{$2\Delta_s$}}
    \Text(75,57)[lb]{\scalebox{0.8}{$x_3$}}
    \Text(33,2)[lb]{\scalebox{0.8}{$x_1$}}
    \Text(110,1)[lb]{\scalebox{0.8}{$x_2$}}
    \Text(153,17)[lb]{\scalebox{1.2}{$=\frac{C_{s^2ss}^{(0)}}{(|x_{13}||x_{23}|)^4}$}}
  \end{picture}
\end{center}
is given by
\begin{equation}
\label{Cs2ss0}
C_{s^2 s s}^{(0)} = 2C_s^2\,,
\end{equation}
where again the factor of 2 comes from the degeneracy of the $s^2$ insertion.

At the next-to-leading order the $\langle s^2 ss\rangle$ three-point function is composed
of the $s$ propagator corrections to the leading-order $\langle s^2 s s\rangle$ diagram,
\begin{equation}
  \begin{picture}(152,55) (36,6)
    \SetWidth{1.0}
    \SetColor{Black}
    \Line[](78,47)(41,11)
    \Line[](79,47)(112,11)
    \GOval(60,29)(7,7)(0){0.882}
    \GOval(95,29)(7,7)(0){0.882}
    \Vertex(41,11){2}
    \Vertex(112,11){2}
    \CBox(76,47)(81,52){Black}{Black}
    \Text(44,32)[lb]{\scalebox{0.8}{$s$}}
    \Text(107,32)[lb]{\scalebox{0.8}{$s$}}
    \Text(75,57)[lb]{\scalebox{0.8}{$x_3$}}
    \Text(33,2)[lb]{\scalebox{0.8}{$x_1$}}
    \Text(110,1)[lb]{\scalebox{0.8}{$x_2$}}
    \Text(153,17)[lb]{\scalebox{1.2}{$=\frac{C_{s^2ss}^{(0)}(1+2A_s)\,\mu^{-4\gamma_s}}{(|x_{13}||x_{23}|)^{4+2\gamma_s}}$}}
  \end{picture}
 \label{first correction to s2ss}
\end{equation}
and three 1PI vertex corrections
\begin{equation}
\scalebox{0.85}{
  \begin{picture}(581,89) (172,24)
    \SetWidth{1.0}
    \SetColor{Black}
    \Line[](246,101)(167,24)
    \Line[](246,101)(316,24)
    \Line[](408,101)(329,24)
    \Line[](576,101)(497,24)
    \Line[](577,101)(647,24)
    \CBox(243,100)(248,105){Black}{Black}
    \CBox(405,100)(410,105){Black}{Black}
    \CBox(574,100)(579,105){Black}{Black}
    \Line[](218,73)(273,73)
    \Line[](190,45)(299,45)
    \Line[](379,73)(434,73)
    \Line[](350,45)(435,45)
    \Line[](479,25)(433,73)
    \Line[](408,101)(434,45)
    \Line[](434,45)(434,73)
    \Vertex(167,23){2}
    \Vertex(316,23){2}
    \Vertex(329,24){2}
    \Vertex(479,24){2}
    \Vertex(498,24){2}
    \Vertex(647,25){2}
    \Line[](554,82)(554,38)
    \Line[](596,82)(597,38)
    \Line[](511,38)(635,38)
    \Vertex(190,45){4}
    \Vertex(218,73){4}
    \Vertex(273,73){4}
    \Vertex(295,45){4}
    \Vertex(379,73){4}
    \Vertex(433,73){4}
    \Vertex(350,45){4}
    \Vertex(433,44){4}
    \Vertex(511,38){4}
    \Vertex(554,82){4}
    \Vertex(554,38){4}
    \Vertex(597,82){4}
    \Vertex(597,38){4}
    \Vertex(633,38){4}
    \Text(245,112)[lb]{\scalebox{0.8}{$x_3$}}
    \Text(405,112)[lb]{\scalebox{0.8}{$x_3$}}
    \Text(576,112)[lb]{\scalebox{0.8}{$x_3$}}
    \Text(165,10)[lb]{\scalebox{0.8}{$x_1$}}
    \Text(310,10)[lb]{\scalebox{0.8}{$x_2$}}
    \Text(334,10)[lb]{\scalebox{0.8}{$x_1$}}
    \Text(480,10)[lb]{\scalebox{0.8}{$x_2$}}
    \Text(496,10)[lb]{\scalebox{0.8}{$x_1$}}
    \Text(648,10)[lb]{\scalebox{0.8}{$x_2$}}
    \Text(323,65)[lb]{\scalebox{0.8}{$+$}}
    \Text(490,62)[lb]{\scalebox{0.8}{$+$}}
    \Text(212,85)[lb]{\scalebox{0.8}{$2\Delta_s$}}
    \Text(264,85)[lb]{\scalebox{0.8}{$2\Delta_s$}}
    \Text(375,87)[lb]{\scalebox{0.8}{$2\Delta_s$}}
    \Text(420,87)[lb]{\scalebox{0.8}{$2\Delta_s$}}
    \Text(550,91)[lb]{\scalebox{0.8}{$2\Delta_s$}}
    \Text(590,91)[lb]{\scalebox{0.8}{$2\Delta_s$}}
    \Text(240,75)[lb]{\scalebox{0.8}{$2\Delta_\phi$}}
    \Text(400,75)[lb]{\scalebox{0.8}{$2\Delta_\phi$}}
    \Text(187,60)[lb]{\scalebox{0.8}{$2\Delta_\phi$}}
    \Text(288,60)[lb]{\scalebox{0.8}{$2\Delta_\phi$}}
    \Text(240,47)[lb]{\scalebox{0.8}{$2\Delta_\phi$}}
    \Text(161,35)[lb]{\scalebox{0.8}{$2\Delta_s$}}
    \Text(309,35)[lb]{\scalebox{0.8}{$2\Delta_s$}}
    \Text(325,35)[lb]{\scalebox{0.8}{$2\Delta_s$}}
    \Text(350,60)[lb]{\scalebox{0.8}{$2\Delta_\phi$}}
    \Text(395,47)[lb]{\scalebox{0.8}{$2\Delta_\phi$}}
    \Text(438,48)[lb]{\scalebox{0.8}{$2\Delta_\phi$}}
    \Text(465,44)[lb]{\scalebox{0.8}{$2\Delta_s$}}
    \Text(514,59)[lb]{\scalebox{0.8}{$2\Delta_\phi$}}
    \Text(559,55)[lb]{\scalebox{0.8}{$2\Delta_\phi$}}
    \Text(582,55)[lb]{\scalebox{0.8}{$2\Delta_\phi$}}
    \Text(620,55)[lb]{\scalebox{0.8}{$2\Delta_\phi$}}
    \Text(572,40)[lb]{\scalebox{0.8}{$2\Delta_s$}}
    \Text(490,35)[lb]{\scalebox{0.8}{$2\Delta_s$}}
    \Text(646,35)[lb]{\scalebox{0.8}{$2\Delta_s$}}
    \Text(528,40)[lb]{\scalebox{0.8}{$2\Delta_\phi$}}
    \Text(612,40)[lb]{\scalebox{0.8}{$2\Delta_\phi$}}
  \end{picture}
}
\label{second correction to s2ss}
\end{equation}
We will denote the total of three diagrams in (\ref{second correction to s2ss}) as
\begin{center}
  \begin{picture}(152,55) (136,6)
    \SetWidth{1.0}
    \SetColor{Black}
    \Line[](78,47)(41,11)
    \Line[](79,47)(112,11)
    \Line[](55,25)(99,25)
    \Line[](58,28)(96,28)
    \Line[](61,31)(93,31)
    \Line[](64,34)(90,34)
    \Line[](67,37)(87,37)
    \Line[](70,40)(85,40)
    \Line[](73,43)(82,43)
    \Vertex(41,11){2}
    \Vertex(112,11){2}
    \CBox(76,47)(81,52){Black}{Black}
    \Text(75,57)[lb]{\scalebox{0.8}{$x_3$}}
    \Text(33,2)[lb]{\scalebox{0.8}{$x_1$}}
    \Text(110,1)[lb]{\scalebox{0.8}{$x_2$}}
    \Text(153,17)[lb]{\scalebox{1.2}{$=\frac{C_{s^2ss}^{(0)}(1+\delta C_{s^2ss}-2A_s)
    \,\mu^{2\gamma_s-\gamma_{s^2}}}{(|x_{13}||x_{23}|)^{4+\gamma_{s^2}
    -2\gamma_s}|x_{12}|^{2\gamma_s-\gamma_{s^2}}}-\frac{C_{s^2ss}^{(0)}}{(|x_{13}||x_{23}|)^4}$}}
  \end{picture}
\end{center}
where the amplitude and the exponents have been determined
from the requirement that the total adds up to (\ref{s2ss general}),
and we also subtracted the leading order $\langle s^2 s s\rangle$ three-point function
to ensure that the result is purely of the next-to-leading order in $1/N$.
We can also re-write this relation as
\begin{equation}
\label{s2ss corrections}
  \begin{picture}(182,55) (136,6)
    \SetWidth{1.0}
    \SetColor{Black}
    \Line[](78,47)(41,11)
    \Line[](79,47)(112,11)
    \Line[](55,25)(99,25)
    \Line[](58,28)(96,28)
    \Line[](61,31)(93,31)
    \Line[](64,34)(90,34)
    \Line[](67,37)(87,37)
    \Line[](70,40)(85,40)
    \Line[](73,43)(82,43)
    \Vertex(41,11){2}
    \Vertex(112,11){2}
    \CBox(76,47)(81,52){Black}{Black}
    \Text(123,17)[lb]{\scalebox{0.8}{$=C_{s^2ss}^{(0)}(1+\delta C_{s^2ss}-2A_s)\times$}}
    \Line[](278,47)(241,11)
    \Line[](279,47)(312,11)
    \Line[](241,11)(312,11)
    \Text(261,0)[lb]{\scalebox{0.8}{$2\gamma_s-\gamma_{s^2}$}}
    \Text(231,30)[lb]{\scalebox{0.8}{$4+\gamma_{s^2}$}}
    \Text(300,30)[lb]{\scalebox{0.8}{$4+\gamma_{s^2}$}}
    \Vertex(241,11){2}
    \Vertex(312,11){2}
    \CBox(276,47)(281,52){Black}{Black}
    \Text(325,17)[lb]{\scalebox{0.8}{$-$}}
    \Line[](378,47)(341,11)
    \Line[](379,47)(412,11)
    \Vertex(341,11){2}
    \Vertex(412,11){2}
    \CBox(376,47)(381,52){Black}{Black}
    \COval(360,29)(7,7)(0){Black}{Black}
    \COval(395,29)(7,7)(0){Black}{Black}
    \Text(346,30)[lb]{\scalebox{0.8}{$s$}}
    \Text(405,30)[lb]{\scalebox{0.8}{$s$}}
  \end{picture}
\end{equation}
We will use the Feynman rule corresponding to the dressed $s^2$ propagator
\begin{center}
  \begin{picture}(147,20) (23,0)
    \SetWidth{1.0}
    \SetColor{Black}
    \Line[](31,5)(112,5)
    \GOval(73,5)(7,7)(0){0.882}
    \Vertex(31,5){2}
    \Vertex(112,5){2}
    \Text(72,15)[lb]{\scalebox{0.8}{$s^2$}}
    \Text(145,-5)[lb]{\scalebox{1}{$=\frac{C_{s^2}\,(1+A_{s^2})\,\mu^{-2\gamma_{s^2}}}{|x|^{2(\Delta_{s^2}+\gamma_{s^2})}}$}}
    \Text(20,4)[lb]{\scalebox{0.8}{$0$}}
    \Text(122,4)[lb]{\scalebox{0.8}{$x$}}
  \end{picture}
\end{center}

\subsection{$s^2 s s$ conformal triangle}

Consider the $s^2 s s$ conformal triangle:
\begin{center}
  \begin{picture}(236,72) (12,17)
    \SetWidth{1.0}
    \SetColor{Black}
    \Line[](144,32)(191,32)
    \Line[](166,69)(145,33)
    \Line[](167,69)(192,33)
    \Line[](143,31)(125,20)
    \Line[](192,32)(209,20)
    \Line[](166,71)(166,88)
    \Line[](42,85)(42,42)
    \Line[](42,42)(73,19)
    \Line[](42,42)(13,18)
    \GOval(42,42)(7,7)(0){0.882}
    \Vertex(166,68){4}
    \Vertex(145,32){4}
    \Vertex(191,32){4}
    \Text(80,45)[lb]{\scalebox{1}{$=-2Z_{s^2 s s}\times$}}
    \Text(186,50)[lb]{\scalebox{0.8}{$2a$}}
    \Text(143,50)[lb]{\scalebox{0.8}{$2a$}}
    \Text(166,23)[lb]{\scalebox{0.8}{$2b$}}
    \Text(47,71)[lb]{\scalebox{0.8}{$s^2$}}
    \Text(10,25)[lb]{\scalebox{0.8}{$s$}}
    \Text(70,25)[lb]{\scalebox{0.8}{$s$}}
    \Text(173,88)[lb]{\scalebox{0.8}{$s^2$}}
    \Text(119,25)[lb]{\scalebox{0.8}{$s$}}
    \Text(213,25)[lb]{\scalebox{0.8}{$s$}}
  \end{picture}
\end{center}
Here we defined
\begin{equation}
a = \frac{d-\gamma_{s^2}}{2}-2\,,\qquad  b= \frac{d+\gamma_{s^2}}{2} - \gamma_s\,.
\end{equation}
We can expand the amplitude of the conformal triangle in $1/N$ as
\begin{equation}
Z_{s^2 s s} = Z_{s^2 s s}^{(0)} (1+\delta Z_{s^2 ss})\,.
\end{equation}
The conformal triangle is defined so that when we attach full propagators of the $s^2$ and $s$
and integrate over the three internal vertices, we obtain the three-point function:
\begin{center}
  \begin{picture}(236,72) (12,17)
    \SetWidth{1.0}
    \SetColor{Black}
    \Line[](42,85)(42,42)
    \Line[](42,42)(73,18)
    \Line[](42,42)(13,18)
    \GOval(42,42)(7,7)(0){0.882}
    \GOval(42,65)(7,7)(0){0.882}
    \GOval(57,30)(7,7)(0){0.882}
    \GOval(27,30)(7,7)(0){0.882}
    \Vertex(42,85){2}
    \Vertex(13,18){2}
    \Vertex(73,18){2}
    \Text(105,45)[lb]{\scalebox{1.2}{$=\frac{C_{s^2ss}^{(0)}(1+\delta C_{s^2ss})
    \,\mu^{-\gamma_{s^2}-2\gamma_s}}{(|x_{13}||x_{23}|)^{4+\gamma_{s^2}}
    |x_{12}|^{2\gamma_s-\gamma_{s^2}}}$}}
    \Text(54,64)[lb]{\scalebox{0.8}{$s^2$}}
    \Text(10,32)[lb]{\scalebox{0.8}{$s$}}
    \Text(70,32)[lb]{\scalebox{0.8}{$s$}}
    \Text(50,84)[lb]{\scalebox{0.8}{$x_3$}}
    \Text(0,12)[lb]{\scalebox{0.8}{$x_1$}}
    \Text(80,12)[lb]{\scalebox{0.8}{$x_2$}}
  \end{picture}
\end{center}
Expanding the integral over the three vertices of the conformal triangle as
\begin{align}
&U\left(4+\gamma_{s^2},\frac{d-\gamma_{s^2}}{2}-2,\frac{d-\gamma_{s^2}}{2}-2\right)
U\left(2+\frac{\gamma_{s^2}}{2},2+\gamma_s,d-4-\gamma_s-\frac{\gamma_{s^2}}{2}\right)\notag\\
&\times U\left(\frac{d+\gamma_{s^2}}{2}-\gamma_s,\frac{d-\gamma_{s^2}}{2}-2,2+\gamma_s\right)
 = u_{s^2 s s}^{(0)}\,\left(1+R_{s^2 ss}+{\cal O}\left(\frac{1}{N^2}\right)\right)\,,
\end{align}
where
\begin{align}
u_{s^2 s s}^{(0)} &=-N\frac{\pi ^{\frac{3 d}{2}+2} \csc ^2\left(\frac{\pi  d}{2}\right) \Gamma \left(\frac{d}{2}+1\right)}{3 (d-8) (d-3) \Gamma (d-4) \Gamma (d+1)}\,,\\
R_{s^2 s s} &=\frac{1}{N}\frac{d \Gamma (d)}{6 \pi  (d-8) (d-6) (d-4) (d-2)  \Gamma \left(\frac{d}{2}+1\right)^2}\\
&\times \left(\left(-3 (d-8) (d-6) (d-4) (d-2) ((d-7) d+8) H_{d-5}\right.\right.\notag\\
&+\left.\left. d (d (d (d (d (5 d-143)+1550)-8252)+23096)-32768)+19200\right) \sin \left(\frac{\pi  d}{2}\right)\right.\notag\\
&-\left. 3 \pi  (d-8) (d-6) (d-4) (d-2) ((d-7) d+8) \cos \left(\frac{\pi  d}{2}\right)\right)\,.\notag
\end{align}
We then obtain
\begin{align}
\label{Zs2ss0}
Z_{s^2 s s}^{(0)} &= -\frac{C_{s^2ss}^{(0)}}{2C_{s^2}C_s^2u_{s^2 s s}^{(0)}}\,,\\
\delta Z_{s^2 s s} &= \delta C_{s^2 s s}-A_{s^2}-2A_s-R_{s^2 s s}\,.
\label{relation between delta Cs2ss and delta Zs2ss}
\end{align}

Up to the next-to-leading order in $1/N$, the $\langle s^2 s^2\rangle$ propagator is given by
\begin{center}
  \begin{picture}(368,60) (67,10)
    \SetWidth{1.0}
    \SetColor{Black}
    \CBox(45,35)(50,40){Black}{Black}
    \Arc[clock](107,-32.976)(91.978,129.496,50.504)
    \Arc[](107,117.5)(98.501,-127.161,-52.839)
    \GOval(108,58)(7,7)(0){0.882}
    \GOval(108,20)(7,7)(0){0.882}
    \Arc[clock](287.5,-32.976)(91.978,129.496,50.504)
    \Arc[](287.5,117.5)(98.501,-127.161,-52.839)
    \CBox(165,35)(170,40){Black}{Black}
    \CBox(224,35)(229,40){Black}{Black}
    \CBox(345,35)(350,40){Black}{Black}
    \Text(37,24)[lb]{\scalebox{0.8}{$0$}}
    \Text(176,24)[lb]{\scalebox{0.8}{$x$}}
    \Text(200,36)[lb]{\scalebox{0.8}{$+$}}
    \Text(90,67)[lb]{\scalebox{0.8}{$2(\Delta_s+\gamma_s)$}}
    \Text(90,2)[lb]{\scalebox{0.8}{$2(\Delta_s+\gamma_s)$}}
    \Text(370,26)[lb]{\scalebox{1.2}{$=\frac{C_{s^2}(1+A_{s^2})\mu^{-2\gamma_{s^2}}}{|x|^{2(\Delta_{s^2}+\gamma_{s^2})}}$}}
    \Text(223,24)[lb]{\scalebox{0.8}{$0$}}
    \Text(355,24)[lb]{\scalebox{0.8}{$x$}}
    \Line[](234,35)(234,42)
    \Line[](237,33)(237,44)
    \Line[](240,31)(240,46)
    \Line[](243,29)(243,48)
    \Line[](246,28)(246,49)
    \Line[](249,26)(249,51)
  \end{picture}
\end{center}

Here the second term stands for dressing one of the $s^2 s s$
sub-diagrams of the leading-order $\langle s^2 s^2\rangle$ diagram, i.e., 
incorporating three 1PI diagrams introduced in (\ref{second correction to s2ss}).
In fact, using the $s^2 s s$ conformal triangle we can re-write the total of the diagrams 
contributing to $\langle s^2 s^2\rangle$ up to the next-to-leading order as\footnote{Such a method of calculation of conformal triangles was first proposed in \cite{Goykhman:2020tsk}, where it was applied to determine the $s^2 s s$ conformal triangle in the Gross-Neveu model.}
\begin{center}
  \begin{picture}(486,118) (35,6)
    \SetWidth{1.0}
    \SetColor{Black}
    \Line[](61,57)(108,57)
    \Line[](109,57)(150,99)
    \Line[](152,100)(224,100)
    \Line[](111,57)(150,15)
    \Line[](151,15)(223,15)
    \Line[](152,100)(152,15)
    \Line[](222,100)(223,14)
    \Line[](223,100)(265,59)
    \Line[](265,57)(225,16)
    \Line[](266,57)(313,57)
    \CBox(58,55)(63,60){Black}{Black}
    \CBox(313,55)(318,60){Black}{Black}
    \Line[](111,57)(150,15)
    \Vertex(111,57){4}
    \Vertex(152,99){4}
    \Vertex(151,14){4}
    \Vertex(222,99){4}
    \Vertex(224,14){4}
    \Vertex(266,56){4}
    \Text(71,60)[lb]{\scalebox{0.8}{$8+2\gamma_{s^2}$}}
    \Text(84,79)[lb]{\scalebox{0.8}{$d-4-\gamma_{s^2}$}}
    \Text(84,31)[lb]{\scalebox{0.8}{$d-4-\gamma_{s^2}$}}
    \Text(155,35)[lb]{\scalebox{0.8}{$d+\gamma_{s^2}-2\gamma_s$}}
    \Text(168,80)[lb]{\scalebox{0.8}{$d+\gamma_{s^2}-2\gamma_s$}}
    \Text(168,103)[lb]{\scalebox{0.8}{$4+2\gamma_s+\delta$}}
    \Text(168,4)[lb]{\scalebox{0.8}{$4+2\gamma_s+\delta$}}
    \Text(248,82)[lb]{\scalebox{0.8}{$d-4-\gamma_{s^2}$}}
    \Text(248,31)[lb]{\scalebox{0.8}{$d-4-\gamma_{s^2}$}}
    \Text(276,60)[lb]{\scalebox{0.8}{$8+2\gamma_{s^2}$}}
    \Text(32,51)[lb]{\scalebox{1}{$\frac{1}{2}\times$}}
    \CBox(364,55)(369,60){Black}{Black}
    \CBox(478,55)(483,60){Black}{Black}
    \Arc[clock](424.152,-13.545)(89.545,128.018,53.033)
    \Arc[](423.5,130.444)(91.457,-126.578,-53.422)
    \GOval(425,76)(11,12)(0){0.882}
    \GOval(425,39)(11,12)(0){0.882}
    \Text(333,51)[lb]{$+\frac{1}{2}\times$}
  \end{picture}
\end{center}
Notice that here the first diagram in fact contains two $s^2 s s$ conformal triangles. To compensate for
such a double counting, however, we multiplied it by the factor of $1/2$.
This diagram
already contains the leading order  $\langle s^2 s^2\rangle$ diagram, as well as its $1/N$ corrections
obtained by dressing of the internal $s$ lines. However, since it
is multiplied by the factor of $1/2$, we need to add another one-half of the leading diagram with
the corrected propagators.
Finally, notice that the first diagram
is divergent. To regularize it, we introduced a small shift $\delta$ to the dressed internal $s$ lines \cite{Vasiliev:1975mq}.

Contribution of the second diagram is given by
\begin{equation}
\label{s2s2 first contribution}
\langle s^2(x)s^2(0)\rangle \supset C_{s^2}\,\frac{1}{|x|^8}\,
\left(\frac{1}{2}+A_s-2\gamma_{s}\log(\mu |x|)\right)
\end{equation}
while contribution of the first diagram is
\begin{align}
\langle s^2(x)s^2(0)\rangle &\supset\frac{1}{2}(C_{s^2}(1+A_{s^2}) C_s (1+A_s)
(-2)Z_{s^2 s s}^{(0)}(1+\delta Z_{s^2 s s}))^2\, V(\delta)\,,
\end{align}
where $V(\delta)$ is obtained by integrating over the internal vertices of the diagram.
To find the latter we first integrate over the left-most and the right-most vertices, resulting in
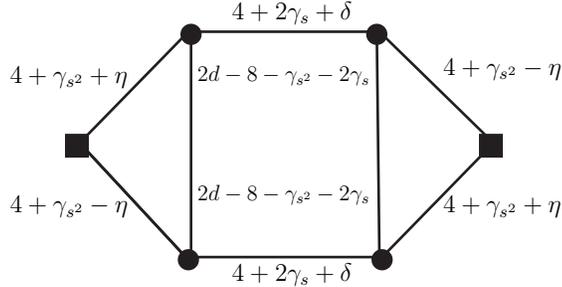
\begin{center}
  \begin{picture}(486,100) (35,6)
    \SetWidth{1.0}
    \SetColor{Black}
    \Line[](109,57)(150,99)
    \Line[](152,100)(224,100)
    \Line[](111,57)(150,15)
    \Line[](151,15)(223,15)
    \Line[](152,100)(152,15)
    \Line[](222,100)(223,14)
    \Line[](223,100)(265,59)
    \Line[](265,57)(225,16)
    \CBox(105,53)(113,61){Black}{Black}
    \CBox(261,53)(269,61){Black}{Black}
    \Line[](111,57)(150,15)
    \Vertex(152,99){4}
    \Vertex(151,14){4}
    \Vertex(222,99){4}
    \Vertex(224,14){4}
    \Text(84,79)[lb]{\scalebox{0.8}{$4+\gamma_{s^2}+\eta$}}
    \Text(84,30)[lb]{\scalebox{0.8}{$4+\gamma_{s^2}-\eta$}}
    \Text(155,35)[lb]{\scalebox{0.7}{$2d-8-\gamma_{s^2}-2\gamma_s$}}
    \Text(155,80)[lb]{\scalebox{0.7}{$2d-8-\gamma_{s^2}-2\gamma_s$}}
    \Text(168,103)[lb]{\scalebox{0.8}{$4+2\gamma_s+\delta$}}
    \Text(168,4)[lb]{\scalebox{0.8}{$4+2\gamma_s+\delta$}}
    \Text(248,82)[lb]{\scalebox{0.8}{$4+\gamma_{s^2}-\eta$}}
    \Text(248,30)[lb]{\scalebox{0.8}{$4+\gamma_{s^2}+\eta$}}
  \end{picture}
\end{center}
Here we have introduced an auxiliary parameter $\eta$ \cite{Vasiliev:1981yc,Vasiliev:1981dg,Gubser:2017vgc} shifting exponents of some
of the lines. One can easily see that the diagram is an even function of $\eta$,\footnote{One can
see this by noticing that $\eta\rightarrow-\eta$ can be undone by swapping vertices of integration
related by mirror reflection in the horizontal axes.}
and as a result choosing $\eta = {\cal O}(\delta)$ will not affect the value of the diagram
in the limit $\delta\rightarrow 0$. We will take advantage of this fact by setting $\eta = \delta$,
which will render two of the vertices unique. Integrating over those vertices we obtain the diagram:
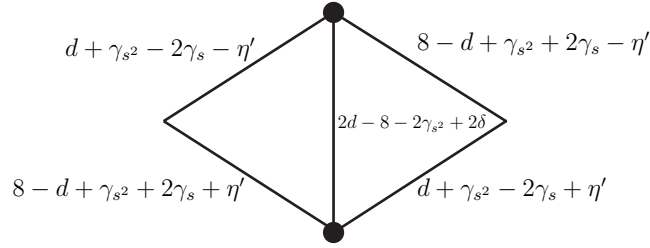
\begin{center}
  \begin{picture}(166,94) (17,-17)
    \SetWidth{1.0}
    \SetColor{Black}
    \Line[](113,70)(113,-12)
    \Line[](113,70)(49,29)
    \Line[](49,29)(113,-12)
    \Line[](113,-13)(178,29)
    \Line[](178,29)(113,70)
    \Text(-8,-1)[lb]{\scalebox{0.8}{$8-d+\gamma_{s^2}+2\gamma_s+\eta'$}}
    \Text(12,52)[lb]{\scalebox{0.8}{$d+\gamma_{s^2}-2\gamma_s-\eta'$}}
    \Text(115,25)[lb]{\scalebox{0.6}{$2d-8-2\gamma_{s^2}+2\delta$}}
    \Text(145,54)[lb]{\scalebox{0.8}{$8-d+\gamma_{s^2}+2\gamma_s-\eta'$}}
    \Text(145,-1)[lb]{\scalebox{0.8}{$d+\gamma_{s^2}-2\gamma_s+\eta'$}}
    \Vertex(113,-13){4}
    \Vertex(113,70){4}
  \end{picture}
\end{center}
Here we introduced yet another auxiliary shift $\eta'$, such that the resulting
diagram is an even function of $\eta'$.\footnote{This can be seen
by renaming the vertices of integration $x_{1,2}$ as $x_1\rightarrow x-x_2$,
$x_2\rightarrow x-x_1$. We refer the reader to \cite{Gubser:2017vgc} for the
detailed explanation of this method of calculating similar diagrams.} Consequently
choosing $\eta' = \delta$ we will not change the value of the diagram
in the $\delta\rightarrow 0$ limit, while this will make the topmost vertex
unique. Completing the last two integrals we obtain for the total:
\begin{align}
V(\delta) &{=} \frac{1}{2}U\left(4{+}\gamma_{s^2},\frac{d{-}\gamma_{s^2}}{2}{-}2,\frac{d{-}\gamma_{s^2}}{2}{-}2\right)^2
U\left(2{+}\gamma_s{+}\frac{\delta}{2},d{-}4{-}\gamma_s{-}\frac{\gamma_{s^2}}{2},
2{+}\frac{\gamma_{s^2}{-}\delta}{2}\right)^2\\
&\times U\left(d{-}4{-}\gamma_{s^2}{+}\delta,\frac{d{+}\gamma_{s^2}{-}\delta}{2}
{-}\gamma_s,\frac{\gamma_{s^2}{-}d{-}\delta}{2}{+}4{+}\gamma_s\right)
U\left(\frac{d}{2}{+}\delta,\frac{d}{2}{+}\delta,{-}2\delta\right)\frac{\mu^{-2\delta}}{|x|^{8+2\gamma_{s^2}+2\delta}}\,,\notag
\end{align}
where $1/2$ is the symmetry factor of the diagram. Expanding the product of the $U$ functions 
around $\delta = 0$ and $N=\infty$ we obtain
\begin{align}
V(\delta) &= v_0\left(1+\frac{\gamma_{s^2}-2\gamma_s}{\delta}
+\delta v\right)\,\frac{\mu^{-2\delta}}{|x|^{8+2\gamma_{s^2}+2\delta}}\\
&=v_0\,\left(1+\delta v+(4\gamma_s-4\gamma_{s^2})\,\log(\mu |x|)\right)\,\frac{1}{|x|^{8}}\,,\notag
\end{align}
where we subtracted divergent part using $s^2ss$ counterterm, and\footnote{Here $\gamma$ is Euler's constant.}
\begin{align}
v_0&=\frac{C_{s^2}}{\left(C_{s^2}\,C_s\,(-2)Z_{s^2 ss}^{(0)}\right)^2}\,,\\
\delta v&=   2   \gamma_s (\gamma-1)  +  \gamma_{s^2}-\frac{8 \gamma_{s^2}}{3}+\frac{2 (2 \gamma_s +\gamma_{s^2})}{d-8}-\frac{4 (d-7) (2 \gamma_s -\gamma_{s^2})}{(d-8) (d-6)}\notag\\
&+\pi  (2 \gamma_s +\gamma_{s^2}) \cot \left(\frac{\pi  d}{2}\right)+(2 \gamma_s +\gamma_{s^2}) \psi ^{(0)}(d-4)\,.
\end{align}
The corresponding contribution to the two-point function is then
\begin{equation}
\label{s2s2 second contribution}
\langle s^2(x)s^2(0)\rangle \supset C_{s^2}\,\left(\frac{1}{2}+A_{s^2}+A_s+\delta Z_{s^2 s s}+\frac{\delta v}{2}
+(2\gamma_{s}-2\gamma_{s^2})\,\log(\mu |x|)\right)\,\frac{1}{|x|^8}\,,
\end{equation}
Combining (\ref{s2s2 first contribution}), (\ref{s2s2 second contribution}) we obtain
\begin{equation}
\langle s^2(x)s^2(0)\rangle = C_{s^2}\,\left(1+A_{s^2}+2A_s+\delta Z_{s^2 s s}+\frac{\delta v}{2}\right)\,\frac{1}{|x|^{8+2\gamma_{s^2}}}\,,
\end{equation}
Consequently
\begin{equation}
\boxed{
\delta Z_{s^2 s s} = - 2A_s - \frac{\delta v}{2}
}
\end{equation}
For the purpose of calculating correction $\delta C_{s^2 s s}$ to the amplitude 
of the three-point function (\ref{s2ss general}), we can now use eq. (\ref{relation between delta Cs2ss and delta Zs2ss}).
Expanding around $d=3$, we notice that the singular parts of two terms $\delta Z_{s^2 s s}$ and
$R_{s^2ss}$ contributing to $\delta C_{s^2 s s}$ cancel each other out:
\begin{align}
\delta Z_{s^2 s s} &= -\frac{1}{N}\frac{64}{3\pi^2 (d-3)}+{\cal O}((d-3)^0)\,,\\
R_{s^2ss} &= \frac{1}{N}\frac{64}{3\pi^2 (d-3)}+{\cal O}((d-3)^0)\,.
\end{align}

Below in section~\ref{sec:<s2ss>} we will calculate the value of $\delta C_{s^2 s s}$ in $d=3$.
Using (\ref{relation between delta Cs2ss and delta Zs2ss}), we then can solve for the value
of the $A_{s^2}$ in $3d$,
\begin{equation}
\label{As2 general}
A_{s^2} = \delta C_{s^2 s s}-\delta Z_{s^2 s s}-2A_s-R_{s^2 s s}\,.
\end{equation}

\subsection{$\langle s^2 s s\rangle$ in $3d$}
\label{sec:<s2ss>}

In this subsection we will calculate the $\langle s^2 s s\rangle$ three-point function in $d=3$
dimensions. The key simplification of considering specifically the three-dimensional
case is that in $3d$ the third 1PI vertex correction diagram in (\ref{second correction to s2ss}) does not contribute
to the three-point function. In fact, in three dimensions it exhibits a vanishing contribution both to the anomalous
dimensions exponents of the three-point function (\ref{s2ss general}), as well as to its overall amplitude.

We have established this as follows. First, one can easily extract
only divergent contributions of the diagrams (\ref{second correction to s2ss}) in any dimension $d$.\footnote{Since these are divergent
diagrams, ordinarily one proceeds by regularizing them via a small correction $\delta$ added to the exponent
of the internal $s$ field propagators \cite{Vasiliev:1975mq}. However, for the purpose of extracting the singular contributions / anomalous dimensions
only, a simpler calculation can be performed, see, e.g., \cite{Chai:2020hnu,Goykhman:2020ffn}.
In such an approach one carries out all of the integrals explicitly, and replaces the logarithmically divergent integral
with $S_d\log \mu$, where $S_d=2\pi^{d/2}/\Gamma(d/2)$ is the surface area of $d-1$-dimensional sphere.
The total of the anomalous dimensions exponents of the considered diagram is then read off from the coefficient
in front of the $\log \mu$ term.}
Setting then $d=3$ one can see that the third diagram in (\ref{second correction to s2ss}) does not contribute
any divergence in three dimensions. Independently, one can verify that
 the contributions of the diagram (\ref{first correction to s2ss}) and the first two diagrams in
 (\ref{second correction to s2ss})
 to the total anomalous
dimensions, entirely account for the anomalous dimensions structure of the three-point function (\ref{s2ss general}). With this result in mind, we conclude that the
third diagram in (\ref{second correction to s2ss}) is finite in $3d$, and proceed to its evaluation without the need to introduce regulators:
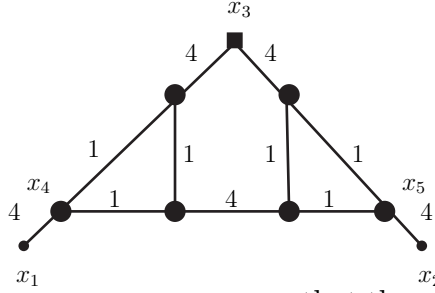
\begin{center}
\scalebox{1}{
  \begin{picture}(381,89) (372,24)
    \SetWidth{1.0}
    \SetColor{Black}
    \Line[](576,101)(497,25)
    \Line[](577,101)(647,25)
    \CBox(574,100)(579,105){Black}{Black}
    \Vertex(497,25){2}
    \Vertex(647,25){2}
    \Line[](554,82)(554,38)
    \Line[](596,82)(597,38)
    \Line[](511,38)(635,38)
    \Vertex(511,38){4}
    \Vertex(554,82){4}
    \Vertex(554,38){4}
    \Vertex(597,82){4}
    \Vertex(597,38){4}
    \Vertex(633,38){4}
    \Text(500,45)[lb]{\scalebox{0.8}{$x_4$}}
    \Text(642,45)[lb]{\scalebox{0.8}{$x_5$}}
    \Text(523,59)[lb]{\scalebox{0.8}{$1$}}
    \Text(559,57)[lb]{\scalebox{0.8}{$1$}}
    \Text(590,57)[lb]{\scalebox{0.8}{$1$}}
    \Text(623,57)[lb]{\scalebox{0.8}{$1$}}
    \Text(590,95)[lb]{\scalebox{0.8}{$4$}}
    \Text(560,95)[lb]{\scalebox{0.8}{$4$}}
    \Text(575,40)[lb]{\scalebox{0.8}{$4$}}
    \Text(493,35)[lb]{\scalebox{0.8}{$4$}}
    \Text(649,35)[lb]{\scalebox{0.8}{$4$}}
    \Text(531,40)[lb]{\scalebox{0.8}{$1$}}
    \Text(612,40)[lb]{\scalebox{0.8}{$1$}}
    \Text(576,112)[lb]{\scalebox{0.8}{$x_3$}}
    \Text(496,10)[lb]{\scalebox{0.8}{$x_1$}}
    \Text(648,10)[lb]{\scalebox{0.8}{$x_2$}}
  \end{picture}
}
\end{center}
 By taking the unique integrals over $x_{4,5}$ one can see that the
 resulting diagram becomes proportional to $\delta^{(3)}(x_{13})\,\delta^{(3)}(x_{23})$\footnote{One can
 see that using the inverse propagator relation
 \begin{equation}
 \int d^dx_3\frac{1}{|x_{13}|^{2\Delta}|x_{23}|^{2(d-\Delta)}} = \pi^d A(\Delta)A(d-\Delta)\delta^{(d)}(x_{12})\,,
 \end{equation}
applied for $d=3$, $\Delta=1$.},
 and is therefore identically zero for the three-point function, that is defined
 for non-coincident points only.
 
 We proceed to the calculation of the contributions of the diagram (\ref{first correction to s2ss})
 and the first two diagrams in (\ref{second correction to s2ss}) to the $\langle s^2 s s\rangle$.
 We performed this calculation in any $d$, so we keep the dimension general, and set $d=3$ in the very end.
 We have the following equation:
 \begin{equation}
 \label{s2ss decomposition}
 (\ref{s2ss general}) = (\ref{first correction to s2ss}) + (\ref{second correction to s2ss})\,.
 \end{equation}
 Following \cite{Goykhman:2020tsk} we replace the $s^2 s s$ sub-diagram in the first
 two diagrams in (\ref{second correction to s2ss}) with the $s^2s s$ conformal
triangle. Such a procedure creates two internal $s$ propagators, which we regularize
by adding a small shift $\delta$ to their exponent:
\begin{equation}
  \begin{picture}(494,119) (88,20)
    \SetWidth{1.0}
    \SetColor{Black}
    \CBox(163,124)(170,131){Black}{Black}
    \Text(163,137)[lb]{\scalebox{0.8}{$x_3$}}
    \Line[](126,24)(166,95)
    \Line[](166,95)(208,24)
    \Line[](150,66)(184,66)
    \Line[](201,38)(134,38)
    \Vertex(134,38){4}
    \Vertex(201,38){4}
    \Vertex(150,66){4}
    \Vertex(184,66){4}
    \Vertex(126,24){2}
    \Vertex(208,24){2}
    \Line[](166,125)(166,94)
    \GOval(166,94)(8,8)(0){0.882}
    \Text(120,77)[lb]{\scalebox{0.8}{$2\Delta_s+\delta$}}
    \Text(187,77)[lb]{\scalebox{0.8}{$2\Delta_s+\delta$}}
    \Line[](275,66)(311,66)
    \Vertex(311,66){4}
    \Line[](251,24)(275,66)
    \Vertex(275,66){4}
    \Line[](335,24)(311,66)
    \Line[](310,38)(260,38)
    \Vertex(260,38){4}
    \Line[](311,66)(310,38)
    \Vertex(310,38){4}
    \Line[](310,38)(350,95)
    \Line[](275,66)(350,95)
    \Vertex(275,66){4}
    \CBox(386,124)(393,131){Black}{Black}
    \Vertex(251,24){2}
    \Vertex(335,24){2}
    \Text(386,137)[lb]{\scalebox{0.8}{$x_3$}}
    \Text(121,13)[lb]{\scalebox{0.8}{$x_1$}}
    \Text(243,13)[lb]{\scalebox{0.8}{$x_1$}}
    \Text(211,13)[lb]{\scalebox{0.8}{$x_2$}}
    \Text(337,13)[lb]{\scalebox{0.8}{$x_2$}}
    \Text(160,3)[lb]{\scalebox{1}{$\textbf{(a)}$}}
    \Text(282,3)[lb]{\scalebox{1}{$\textbf{(b)}$}}
    \Text(420,53)[lb]{\scalebox{0.8}{$\subset\langle s(x_3)^2s(x_1)s(x_2)\rangle$}}
    \Text(230,59)[lb]{\scalebox{0.8}{$+$}}
    \Line[](350,95)(390,125)
    \GOval(350,95)(8,8)(0){0.882}
    \Text(290,87)[lb]{\scalebox{0.8}{$2\Delta_s+\delta$}}
    \Text(340,67)[lb]{\scalebox{0.8}{$2\Delta_s+\delta$}}
  \end{picture}
  \label{<s^2 s s> diagrams}
\end{equation}
 
We proceed by integrating both sides of (\ref{s2ss decomposition}) over $x_3$.
Due to (\ref{s2ss general}) on the l.h.s. of (\ref{s2ss decomposition}) we obtain
 \begin{align}
 \label{lhss2ss}
&\textrm{L.H.S.}=C_{s^2 s s}^{(0)}(1+\delta C_{s^2 s s})U\left(2+\frac{\gamma_{s^2}}{2},
 2+\frac{\gamma_{s^2}}{2},d-4-\gamma_{s^2}\right)\,\frac{\mu^{-\gamma_{s^2}-2\gamma_s}}{|x_{12}|^{8-d+2\gamma_s+\gamma_{s^2}}}\notag\\
 &=2C_s^2U(2,2,d-4)(1+\delta C_{s^2 s s}+h_1)\,\frac{\mu^{-\gamma_{s^2}-2\gamma_s}}{|x_{12}|^{8-d+2\gamma_s+\gamma_{s^2}}}\,,\\
 h_1&=-\frac{1}{N}\frac{2^{d+3} (d-1) \sin \left(\frac{\pi  d}{2}\right) \Gamma \left(\frac{d+1}{2}\right) \left(-\frac{2}{d-6}+\pi  \cot \left(\frac{\pi  d}{2}\right)+\psi ^{(0)}(d-4)+\gamma -1\right)}{\pi ^{3/2} (d-2) d \Gamma \left(\frac{d}{2}-2\right)}\,.
 \label{h1}
 \end{align}
 
Next, using (\ref{first correction to s2ss}) we obtain the following contribution due to the
tree-level diagram with dressed $s$ propagators:
\begin{align}
&\int d^dx_3\frac{C_{s^2ss}^{(0)}(1+2A_s)\mu^{-4\gamma_s}}{(|x_{13}||x_{23}|)^{4+2\gamma_s}}
=2C_s^2(1+2A_s)U(2+\gamma_s,2+\gamma_s,d-4-2\gamma_s)\frac{\mu^{-4\gamma_s}}{|x_{12}|^
{8-d+4\gamma_s}}\notag\\
&=2C_s^2(1+2A_s+h_2)U(2,2,d-4)\frac{\mu^{-4\gamma_s}}{|x_{12}|^ {8-d+4\gamma_s}}\,,\label{1s2ss}\\
h_2&=\frac{1}{N}\frac{8 \sin \left(\frac{\pi  d}{2}\right) \Gamma (d) \left((d-6) H_{d-5}-d+\pi  (d-6) \cot \left(\frac{\pi  d}{2}\right)+4\right)}{\pi  (d-6) \Gamma \left(\frac{d}{2}-1\right) \Gamma \left(\frac{d}{2}+1\right)}\,.\label{h2}
\end{align}
 
 Integrating diagrams $\textbf{(a)}$, $\textbf{(b)}$ in figure (\ref{<s^2 s s> diagrams}) over $x_3$ we obtain
 \begin{equation}
 \int d^dx_3\frac{\langle s(x_3)^2s(x_1)s(x_2)\rangle}{2C_s^2 U(2,2,d-4)}\supset
 \left(h_{a,b}-\omega_{a,b}\log(\mu |x_{12}|)\right)\frac{1}{|x_{12}|^{8-d}}\,.
 \end{equation}
 Analogously, integrating the third diagram in (\ref{second correction to s2ss})
 we obtain
 \begin{equation}
 \int d^dx_3\frac{\langle s(x_3)^2s(x_1)s(x_2)\rangle}{2C_s^2 U(2,2,d-4)}\supset
 \left(h_{c}-\omega_{c}\log(\mu |x_{12}|)\right)\frac{1}{|x_{12}|^{8-d}}\,.
 \end{equation}
For our purposes, we only need to know
\begin{equation}
\label{hc in 3d}
h_c(d=3) = 0\,,\qquad \omega_c(d=3) = 0\,,
\end{equation} 
as we established above.
Integrating over vertices of the $s^2ss$ conformal triangle in the 
diagrams $\textbf{(a)}$, $\textbf{(b)}$ in figure (\ref{<s^2 s s> diagrams}), and collecting
the $\delta$-dependent factors, we obtain
\begin{align}
&A\left(2{+}\frac{\delta}{2}\right)A\left(\frac{d{-}\delta}{2}{-}2\right)
U\left(2{+}\frac{\delta}{2},2{+}\frac{\delta}{2},d{-}4{-}\delta\right){=}
A(2)A\left(\frac{d}{2}{-}2\right)U(2,2,d{-}4)(1{+}r\delta)\,,\notag\\
r&=-\frac{2}{d-6}+\pi  \cot \left(\frac{\pi  d}{2}\right)+\psi ^{(0)}(d-4)+\gamma -1\,.
\end{align}
In the process we dropped the $1/N$ corrections due to the
anomalous dimensions, which are sub-leading in $1/N$.\footnote{The $U$ functions generated
due to integrals over the vertices of the $s^2 s s$ conformal triangle, contain the factor $A\left(\frac{d+\gamma_{s^2}}{2}-\gamma_s\right)$. We have expanded it in $1/N$ and kept the leading ${\cal O}(N)$ term.} At the same time,
using (\ref{Zs2ss0}), (\ref{Cs2ss0}), the leading order factors can be assembled into
the leading order amplitude $C_{s^2 s s}^{(0)} = 2C_s^2$.
The factor of $1+r\,\delta$ will be important below for the calculation of the finite
correction $\delta C_{s^2 s s}$ to the amplitude of the three-point function $\langle s^2 s s\rangle$.
Integrating over the remaining four vertices of the $\textbf{(a)}$ in figure (\ref{<s^2 s s> diagrams})
we obtain
\begin{align}
&\int d^dx_3\frac{\langle s(x_3)^2s(x_1)s(x_2)\rangle}{2C_s^2 U(2,2,d-4)}\supset
NC_s^2C_\phi^4(1+r\delta)\left(-\frac{2}{\sqrt{N}}\right)^4
U\left(3+\delta,\frac{d}{2}-1,\frac{d}{2}-2-\delta\right)\notag\\
&U\left(\frac{d}{2}-1,2+\delta,\frac{d}{2}-1-\delta\right)
U\left(\frac{d}{2}+\delta,2,\frac{d}{2}-2-\delta\right)
U(2,2+\delta,d-4-\delta)\frac{\mu^{-2\delta}}{x^{8-d+2\delta}}\,.
\end{align}
Expanding it in $\delta$ we obtain
\begin{align}
\label{omegaa}
\omega_a &= \frac{d-4}{2(d-1)}(2\gamma_s-\gamma_{s^2})\,,\\
h_a&=\frac{1}{N}\frac{4 (d-3) \Gamma \left(4-\frac{d}{2}\right) \Gamma (d-2)\sin ^2\left(\frac{\pi  d}{2}\right)}
{\pi ^2 (d-6)^2 (d-2) \Gamma \left(\frac{d}{2}\right)}\left(
d (5 d-32)-4 \pi  (d-6) (d-2) \cot \left(\frac{\pi  d}{2}\right)\right.\notag\\
&+\left.28-4 (d-6) (d-2) H_{d-5}\right)\,.\label{ha}
\end{align}

Integrating over the remaining four vertices of the $\textbf{(b)}$ in figure (\ref{<s^2 s s> diagrams})
we obtain
\begin{align}
\label{nonplanar s2ss}
&\int d^dx_3\frac{\langle s(x_3)^2s(x_1)s(x_2)\rangle}{2C_s^2 U(2,2,d-4)}{\supset}
\frac{1}{2}NC_s^2C_\phi^4(1{+}r\delta)\left({-}\frac{2}{\sqrt{N}}\right)^4
U\left(\frac{d}{2}{-}1,\frac{d}{2}{-}1,2\right)^2 \frac{\textrm{sk}\left(\frac{d}{2}{+}\delta\right)\mu^{-2\delta}}{|x_{12}|^{8-d+2\delta}}\,,
\end{align}
where $1/2$ is the symmetry factor, and we used the special kite diagram
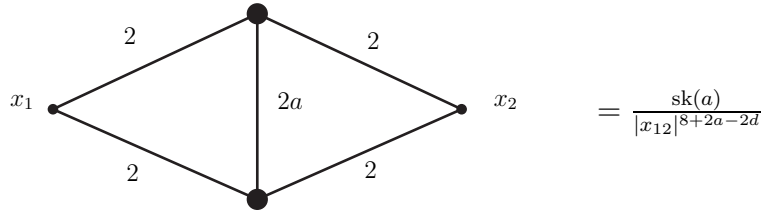
\begin{center}
  \begin{picture}(223,85) (41,-19)
    \SetWidth{1.0}
    \SetColor{Black}
    \Line[](43,20)(120,56)
    \Line[](120,56)(197,20)
    \Line[](197,20)(120,-14)
    \Line[](43,20)(120,-14)
    \Line[](120,56)(120,-14)
    \Vertex(120,56){4}
    \Vertex(120,-14){4}
    \Vertex(43,20){2}
    \Vertex(197,20){2}
    \Text(27,20)[lb]{\scalebox{0.8}{$x_1$}}
    \Text(210,20)[lb]{\scalebox{0.8}{$x_2$}}
    \Text(70,45)[lb]{\scalebox{0.8}{$2$}}
    \Text(162,43)[lb]{\scalebox{0.8}{$2$}}
    \Text(128,20)[lb]{\scalebox{0.8}{$2a$}}
    \Text(71,-7)[lb]{\scalebox{0.8}{$2$}}
    \Text(161,-6)[lb]{\scalebox{0.8}{$2$}}
    \Text(249,10)[lb]{\scalebox{1}{$=\frac{\textrm{sk}(a)}{|x_{12}|^{8+2a-2d}}$}}
  \end{picture}
\end{center}
For general $a$, the value of the special kite diagram can be found in \cite{Kotikov:2000yd}.\footnote{See eq. (22) therein; notice that each integral over internal vertex in \cite{Kotikov:2000yd} is multiplied by $1/(2\pi)^d$, so we need
to multiply that expression by $(2\pi)^{2d}$ to adjust it to our conventions.} However, since we are only interested
in expansion around $\delta=0$ and retaining only singular and finite terms, it is sufficient to use
expression (\ref{selfenergy}) for $\alpha_{1,2,3,4}=1$, $\alpha_5 = d/2+\delta$, and expand around $\delta=0$.
While the 2nd and 3rd diagrams in the r.h.s. of  (\ref{selfenergy})  are straightforward to calculate 
using the propagator merging relation, the first diagram can be calculated by inserting a point into the diagonal
propagator, splitting it into two propagators with the exponents $2d-4$ and $2+2\delta$. Taking the unique
integral we will obtain the diagram equal to $F\left(\frac{d}{2}-1,\frac{d}{2}-1\right)$, where we dropped corrections
linear in $\delta$, since the diagram is finite. Assembling everything together, we obtain
\begin{align}
&\textrm{sk}\left(\frac{d}{2}+\delta\right) = \frac{f_1}{\delta}+f_2+{\cal O}(\delta)\,,\\
f_1 &=\frac{2 (d-6) \pi ^{d+1} \csc \left(\frac{\pi  d}{2}\right)}{(d-2)   \Gamma (d-3)}\,,\\
f_2&{=}\frac{\pi^d}{2}\left(
\frac{4 \pi  \csc \left(\frac{\pi  d}{2}\right)}{(d-2)^2 \Gamma (d-3)}
\left({-}2 (d{-}5) d{+}\pi  (d{-}6) (d{-}2) \cot \left(\frac{\pi  d}{2}\right){-}4
{+}(d{-}6) (d{-}2) H_{d{-}4}\right)\right.\notag\\
&+\left.(d-4) \cos \left(\frac{\pi  d}{2}\right) \Gamma (3-d) \left(\pi ^2-6 \psi ^{(1)}\left(\frac{d}{2}-1\right)\right)\right)\,.
\end{align}
Using it in (\ref{nonplanar s2ss})
and expanding it in $\delta$ we obtain
\begin{align}
\label{omegab}
\omega_b &= \frac{d-6}{2(d-1)}(2\gamma_s-\gamma_{s^2})\,,\\
h_b&=\frac{1}{N}\frac{4^{d-3} \sin \left(\frac{\pi  d}{2}\right) \Gamma \left(\frac{d-1}{2}\right)^2}{\pi ^2 (d-2)^2 \Gamma (d-3) \Gamma (d-1)}\left(-4 (3-d) (d-2)^2 \Gamma (d-3) \left(\gamma  (d-6)-d\right.\right.\notag\\
&+\left.\left.\pi  (d-6) \cot \left(\frac{\pi  d}{2}\right)+(d-6) \psi ^{(0)}(d-4)+4\right)+
\Gamma (d-1)\left(-8 (d-5) d-16\right.\right.\notag\\
&+\left.\left.4 (d-6) (d-2) H_{d-4}+4 \pi  (d-6) (d-2) \cot \left(\frac{\pi  d}{2}\right)\right.\right.\notag\\
&+\left.\left.
\frac{(d-4) (d-2)^2 \sin (\pi  d) \Gamma (3-d) \Gamma (d-3) \left(\pi ^2-6 \psi ^{(1)}\left(\frac{d}{2}-1\right)\right)}{2 \pi }\right)
\right)\,.\label{hb}
\end{align}

Using (\ref{1s2ss}), (\ref{omegaa}), (\ref{omegab}) in $d=3$ we obtain
the total anomalous dimensions term $-(\gamma_{s^2}+2\gamma_s)\log(\mu |x_{12}|)$,
in agreement with (\ref{s2ss decomposition}), (\ref{lhss2ss}). At the same time, using (\ref{s2ss decomposition}),
(\ref{h1}), (\ref{h2}), (\ref{ha}), (\ref{hb}) we obtain
\begin{equation}
\delta C_{s^2 s s} = -h_1 +h_2+2A_s+ h_a+h_b+h_c\,.
\end{equation}
Evaluating in $d=3$ and using (\ref{As2 general}) gives\footnote{Recall that $h_c(d=3)=0$, due to (\ref{hc in 3d}).}
\begin{equation}
\boxed{
A _{s^2}(d=3)=\delta C_{s^2 ss}(d=3)=\frac{1}{N}\,\left(\frac{176}{9\pi^2} - 1\right)
}
\end{equation}

\section{$\langle s^2 s^2 s\rangle$}
\label{s2s2s}

In this section we are going to calculate the $\langle s^2 s^2 s\rangle$
three-point function at the next-to-leading order in the $1/N$ expansion.
Conformal symmetry demands it to have the form
\begin{equation}
\label{s2s2s general}
\langle s^2(x_1) s^2(x_2) s(x_3)\rangle = \frac{C_{s^2 s^2 s}^{(0)}(1+\delta C_{s^2 s^2 s})\,\mu^{-\gamma_s-2\gamma_{s^2}}}{|x_{12}|^{6+2\gamma_{s^2}-\gamma_s}
(|x_{13}||x_{23}|)^{2+\gamma_s}}\,.
\end{equation}
Our goal is to find the amplitude correction $\delta C_{s^2 s^2 s}$.

At the leading ${\cal O}(1/\sqrt{N})$ order in $1/N$ the $\langle s^2 s^2 s\rangle$
three-point function is determined by the diagram
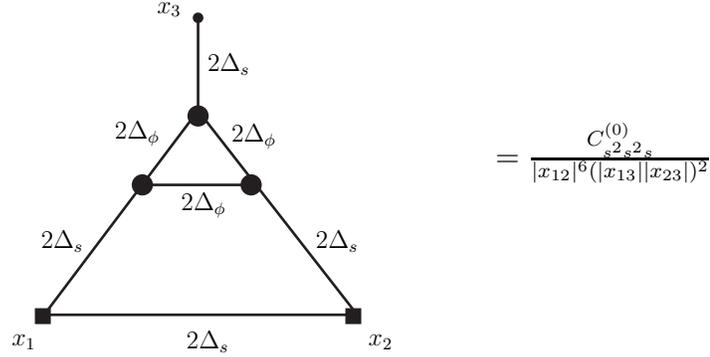
\begin{center}
  \begin{picture}(215,125) (39,7)
    \SetWidth{1.0}
    \SetColor{Black}
    \Line[](106,130)(106,95)
    \Line[](85,67)(128,67)
    \Line[](106,95)(48,18)
    \Line[](106,95)(166,18)
    \Line[](48,18)(166,18)
    \CBox(45,15)(50,20){Black}{Black}
    \CBox(162,15)(167,20){Black}{Black}
    \Vertex(106,93){4}
    \Vertex(85,67){4}
    \Vertex(126,67){4}
    \Vertex(106,130){2}
    \Text(110,109)[lb]{\scalebox{0.8}{$2\Delta_s$}}
    \Text(75,82)[lb]{\scalebox{0.8}{$2\Delta_\phi$}}
    \Text(119,81)[lb]{\scalebox{0.8}{$2\Delta_\phi$}}
    \Text(100,55)[lb]{\scalebox{0.8}{$2\Delta_\phi$}}
    \Text(47,42)[lb]{\scalebox{0.8}{$2\Delta_s$}}
    \Text(151,42)[lb]{\scalebox{0.8}{$2\Delta_s$}}
    \Text(102,4)[lb]{\scalebox{0.8}{$2\Delta_s$}}
    \Text(91,131)[lb]{\scalebox{0.8}{$x_3$}}
    \Text(36,5)[lb]{\scalebox{0.8}{$x_1$}}
    \Text(171,5)[lb]{\scalebox{0.8}{$x_2$}}
    \Text(219,68)[lb]{\scalebox{1}{$=\frac{C_{s^2 s^2 s}^{(0)}}{|x_{12}|^6(|x_{13}||x_{23}|)^2}$}}
  \end{picture}
\end{center}
where the leading order amplitude is given by\footnote{In such diagrams
the factor of $2^2$ is a degeneracy factor due to the composite operators $s^2$.}
\begin{equation}
\boxed{
C_{s^2 s^2 s}^{(0)} = 2^2 C_{sss}^{(0)} C_s
}
\end{equation}
This amplitude vanishes in $3d$, owing to the fact that $C_{sss}^{(0)}(d=3) = 0$.

\subsection*{Next-to-leading order}

We now proceed to calculation of the $\langle s^2 s^2 s\rangle$
three-point function at the next-to-leading order in the $1/N$ expansion.
Let us begin by considering the contributing diagrams which are obtained
by incorporating the $1/N$ corrections to the $s$ propagators and the
and the $sss$ sub-diagram of the leading order $\langle s^2 s^2 s\rangle$ diagram:
\begin{equation}
  \begin{picture}(415,125) (39,7)
    \SetWidth{1.0}
    \SetColor{Black}
    \Line[](106,130)(106,95)
    \Line[](106,95)(48,18)
    \Line[](106,95)(166,18)
    \Line[](48,18)(166,18)
    \Vertex(106,130){2}
    \Text(119,115)[lb]{\scalebox{0.8}{$s$}}
    \Text(62,62)[lb]{\scalebox{0.8}{$s$}}
    \Text(148,62)[lb]{\scalebox{0.8}{$s$}}
    \Text(105,30)[lb]{\scalebox{0.8}{$s$}}
    \Text(91,131)[lb]{\scalebox{0.8}{$x_3$}}
    \Text(36,5)[lb]{\scalebox{0.8}{$x_1$}}
    \Text(171,5)[lb]{\scalebox{0.8}{$x_2$}}
    \GOval(106,94)(7,7)(0){0.882}
    \GOval(106,114)(7,7)(0){0.882}
    \GOval(77,56)(7,7)(0){0.882}
    \GOval(135,56)(7,7)(0){0.882}
    \GOval(107,18)(7,7)(0){0.882}
    \CBox(46,15)(51,20){Black}{Black}
    \CBox(163,15)(168,20){Black}{Black}
    \Text(169,58)[lb]{\scalebox{0.8}{$=(1+4A_s+\delta Z_{sss})\times$}}
    \Line[](306,130)(306,95)
    \Line[](306,95)(248,18)
    \Line[](306,95)(366,18)
    \Line[](248,18)(366,18)
    \Vertex(306,130){2}
    \Text(319,115)[lb]{\scalebox{0.8}{$s$}}
    \Text(262,62)[lb]{\scalebox{0.8}{$s$}}
    \Text(348,62)[lb]{\scalebox{0.8}{$s$}}
    \Text(305,30)[lb]{\scalebox{0.8}{$s$}}
    \Text(291,131)[lb]{\scalebox{0.8}{$x_3$}}
    \Text(236,5)[lb]{\scalebox{0.8}{$x_1$}}
    \Text(371,5)[lb]{\scalebox{0.8}{$x_2$}}
    \COval(306,94)(7,7)(0){Black}{Black}
    \COval(306,114)(7,7)(0){Black}{Black}
    \COval(277,56)(7,7)(0){Black}{Black}
    \COval(335,56)(7,7)(0){Black}{Black}
    \COval(307,18)(7,7)(0){Black}{Black}
    \CBox(246,15)(251,20){Black}{Black}
    \CBox(363,15)(368,20){Black}{Black}
    \Text(370,55)[lb]{\scalebox{0.8}{$=\frac{C_{s^2 s^2 s}^{(0)}(1+v_1)\mu^{-5\gamma_s}}{|x_{12}|^{6+3\gamma_s}
    (|x_{13}||x_{23}|)^{2+\gamma_s}}$}}
  \end{picture}
  \label{total sss correction in s2s2s}
\end{equation}
Here we have used the $sss$ conformal triangle to integrate the diagram
\begin{equation}
  \begin{picture}(415,125) (139,7)
    \SetWidth{1.0}
    \SetColor{Black}
    \Line[](306,130)(306,95)
    \Line[](306,95)(248,18)
    \Line[](306,95)(366,18)
    \Line[](248,18)(366,18)
    \Vertex(306,130){2}
    \Text(319,115)[lb]{\scalebox{0.8}{$s$}}
    \Text(262,62)[lb]{\scalebox{0.8}{$s$}}
    \Text(348,62)[lb]{\scalebox{0.8}{$s$}}
    \Text(305,30)[lb]{\scalebox{0.8}{$s$}}
    \Text(291,131)[lb]{\scalebox{0.8}{$x_3$}}
    \Text(236,5)[lb]{\scalebox{0.8}{$x_1$}}
    \Text(371,5)[lb]{\scalebox{0.8}{$x_2$}}
    \COval(306,94)(7,7)(0){Black}{Black}
    \COval(306,114)(7,7)(0){Black}{Black}
    \COval(277,56)(7,7)(0){Black}{Black}
    \COval(335,56)(7,7)(0){Black}{Black}
    \COval(307,18)(7,7)(0){Black}{Black}
    \CBox(246,15)(251,20){Black}{Black}
    \CBox(363,15)(368,20){Black}{Black}
    \Text(370,55)[lb]{\scalebox{0.8}{$=\frac{C_{s^2 s^2 s}^{(0)}(1+R_{sss})\mu^{-5\gamma_s}}{|x_{12}|^{6+3\gamma_s}
    (|x_{13}||x_{23}|)^{2+\gamma_s}}$}}
  \end{picture}
\label{sss correction in s2s2s}
\end{equation}
obtaining
\begin{equation}
\label{v1}
v_1 = 4A_s+\delta Z_{sss} +R_{sss}=\delta C_{sss} + A_s\,,
\end{equation}
where in the last equality we used (\ref{deltaZsss in terms of deltaCsss}).

Next we consider $1/N$ corrections to the $s^2 s s$ sub-diagrams of the 
the leading order $\langle s^2 s^2 s\rangle$. We will demonstrate below that
these diagrams have the form
\begin{equation}
  \begin{picture}(415,125) (39,7)
    \SetWidth{1.0}
    \SetColor{Black}
    \Line[](106,130)(106,95)
    \Line[](85,67)(128,67)
    \Line[](106,95)(47,18)
    \Line[](106,95)(166,18)
    \Line[](48,18)(166,18)
    \Line[](55,18)(51,24)
    \Line[](58,18)(53,26)
    \Line[](61,18)(55,28)
    \Line[](64,18)(57,30)
    \Line[](67,18)(58,33)
    \Line[](70,18)(60,35)
    \Line[](73,18)(62,37)
    \Line[](158,18)(162,24)
    \Line[](155,18)(160,26)
    \Line[](152,18)(158,28)
    \Line[](149,18)(156,30)
    \Line[](146,18)(154,32)
    \Line[](143,18)(153,35)
    \Line[](140,18)(151,37)
    \CBox(45,15)(50,20){Black}{Black}
    \CBox(162,15)(167,20){Black}{Black}
    \Vertex(106,93){4}
    \Vertex(85,67){4}
    \Vertex(126,67){4}
    \Vertex(106,130){2}
    \Text(110,109)[lb]{\scalebox{0.8}{$2\Delta_s$}}
    \Text(75,82)[lb]{\scalebox{0.8}{$2\Delta_\phi$}}
    \Text(119,81)[lb]{\scalebox{0.8}{$2\Delta_\phi$}}
    \Text(100,55)[lb]{\scalebox{0.8}{$2\Delta_\phi$}}
    \Text(91,131)[lb]{\scalebox{0.8}{$x_3$}}
    \Text(36,5)[lb]{\scalebox{0.8}{$x_1$}}
    \Text(171,5)[lb]{\scalebox{0.8}{$x_2$}}
    \Text(200,68)[lb]{\scalebox{1}{$=\frac{C_{s^2 s^2 s}^{(0)}(1+v_2+v_3)\mu^{4\gamma_s-2\gamma_{s^2}}}{|x_{12}|^{6+2\gamma_{s^2}-4\gamma_s}
    (|x_{13}||x_{23}|)^{2}}-\frac{C_{s^2 s^2 s}^{(0)}}{|x_{12}|^6(|x_{13}||x_{23}|)^2}$}}
  \end{picture}
  \label{s2ss corrections in s2s2s}
\end{equation}
We proceed to calculating diagrams involving corrections to the left-hand and the right-hand $s^2 s s$
sub-diagrams separately. The total of these contributions is additive, due to linearization of the
next-to-leading $1/N$ corrections.
Using (\ref{s2ss corrections}) we can express
\begin{center}
  \begin{picture}(415,125) (39,7)
    \SetWidth{1.0}
    \SetColor{Black}
    \Line[](106,130)(106,95)
    \Line[](85,67)(128,67)
    \Line[](106,95)(47,18)
    \Line[](106,95)(166,18)
    \Line[](48,18)(166,18)
    \Line[](55,18)(51,24)
    \Line[](58,18)(53,26)
    \Line[](61,18)(55,28)
    \Line[](64,18)(57,30)
    \Line[](67,18)(58,33)
    \Line[](70,18)(60,35)
    \Line[](73,18)(62,37)
    \CBox(45,15)(50,20){Black}{Black}
    \CBox(162,15)(167,20){Black}{Black}
    \Vertex(106,93){4}
    \Vertex(85,67){4}
    \Vertex(126,67){4}
    \Vertex(106,130){2}
    \Text(110,109)[lb]{\scalebox{0.8}{$2\Delta_s$}}
    \Text(75,82)[lb]{\scalebox{0.8}{$2\Delta_\phi$}}
    \Text(119,81)[lb]{\scalebox{0.8}{$2\Delta_\phi$}}
    \Text(100,55)[lb]{\scalebox{0.8}{$2\Delta_\phi$}}
    \Text(150,43)[lb]{\scalebox{0.8}{$2\Delta_s$}}
    \Text(91,131)[lb]{\scalebox{0.8}{$x_3$}}
    \Text(36,5)[lb]{\scalebox{0.8}{$x_1$}}
    \Text(171,5)[lb]{\scalebox{0.8}{$x_2$}}
    \Text(145,65)[lb]{\scalebox{0.75}{$=C_{s^2ss}^{(0)}(1+\delta C_{s^2ss}-2A_s)\times $}}
    \Line[](276,130)(276,95)
    \Line[](255,67)(298,67)
    \Line[](276,95)(217,18)
    \Line[](276,95)(336,18)
    \Line[](218,18)(336,18)
    \Line[](255,67)(336,18)
    \Text(255,35)[lb]{\scalebox{0.8}{$2\gamma_s-\gamma_{s^2}$}}
    \Text(205,40)[lb]{\scalebox{0.8}{$4+\gamma_{s^2}$}}
    \Text(267,5)[lb]{\scalebox{0.8}{$4+\gamma_{s^2}$}}
    \CBox(215,15)(220,20){Black}{Black}
    \CBox(332,15)(337,20){Black}{Black}
    \Vertex(276,93){4}
    \Vertex(255,67){4}
    \Vertex(296,67){4}
    \Vertex(276,130){2}
    \Text(280,109)[lb]{\scalebox{0.8}{$2\Delta_s$}}
    \Text(245,82)[lb]{\scalebox{0.8}{$2\Delta_\phi$}}
    \Text(289,81)[lb]{\scalebox{0.8}{$2\Delta_\phi$}}
    \Text(270,70)[lb]{\scalebox{0.8}{$2\Delta_\phi$}}
    \Text(320,43)[lb]{\scalebox{0.8}{$2\Delta_s$}}
    \Text(261,131)[lb]{\scalebox{0.8}{$x_3$}}
    \Text(206,5)[lb]{\scalebox{0.8}{$x_1$}}
    \Text(341,5)[lb]{\scalebox{0.8}{$x_2$}}
    \Text(345,65)[lb]{\scalebox{1}{$-$}}
    \Line[](426,130)(426,95)
    \Line[](405,67)(448,67)
    \Line[](426,95)(367,18)
    \Line[](426,95)(486,18)
    \Line[](368,18)(486,18)
    \CBox(365,15)(370,20){Black}{Black}
    \CBox(482,15)(487,20){Black}{Black}
    \Vertex(426,93){4}
    \Vertex(405,67){4}
    \Vertex(446,67){4}
    \Vertex(426,130){2}
    \Text(430,109)[lb]{\scalebox{0.8}{$2\Delta_s$}}
    \Text(395,82)[lb]{\scalebox{0.8}{$2\Delta_\phi$}}
    \Text(439,81)[lb]{\scalebox{0.8}{$2\Delta_\phi$}}
    \Text(420,70)[lb]{\scalebox{0.8}{$2\Delta_\phi$}}
    \Text(470,43)[lb]{\scalebox{0.8}{$2\Delta_s$}}
    \Text(411,131)[lb]{\scalebox{0.8}{$x_3$}}
    \Text(356,5)[lb]{\scalebox{0.8}{$x_1$}}
    \Text(491,5)[lb]{\scalebox{0.8}{$x_2$}}
    \COval(425,18)(7,7)(0){Black}{Black}
    \COval(385,43)(7,7)(0){Black}{Black}
  \end{picture}
\end{center}
By linearizing over the $1/N$ corrections to the exponents of propagators
of diagrams with identical leading-order skeleton structure, we can further re-write this
expression as
\begin{center}
  \begin{picture}(515,125) (89,7)
    \SetWidth{1.0}
    \SetColor{Black}
    \Text(100,65)[lb]{\scalebox{0.9}{$-\frac{8C_\phi^3}{\sqrt{N}}C_{s^2ss}^{(0)}(1+\delta C_{s^2ss}-2A_s)\times $}}
    \Line[](276,130)(276,95)
    \Line[](255,67)(298,67)
    \Line[](276,95)(217,18)
    \Line[](276,95)(336,18)
    \Line[](218,18)(336,18)
    \Line[](255,67)(336,18)
    \Text(255,35)[lb]{\scalebox{0.8}{$2\gamma_s-\gamma_{s^2}$}}
    \Text(205,40)[lb]{\scalebox{0.8}{$4+\gamma_{s^2}$}}
    \Text(267,5)[lb]{\scalebox{0.8}{$4+\gamma_{s^2}$}}
    \CBox(215,15)(220,20){Black}{Black}
    \CBox(332,15)(337,20){Black}{Black}
    \Vertex(276,93){4}
    \Vertex(255,67){4}
    \Vertex(296,67){4}
    \Vertex(276,130){2}
    \Text(230,82)[lb]{\scalebox{0.6}{$d-2-\gamma_s$}}
    \Text(292,81)[lb]{\scalebox{0.6}{$d-2-\gamma_s$}}
    \Text(263,70)[lb]{\scalebox{0.5}{$d-2-\gamma_s$}}
    \Text(261,131)[lb]{\scalebox{0.8}{$x_3$}}
    \Text(206,5)[lb]{\scalebox{0.8}{$x_1$}}
    \Text(341,5)[lb]{\scalebox{0.8}{$x_2$}}
    \COval(276,113)(7,7)(0){Black}{Black}
    \Text(290,113)[lb]{\scalebox{0.8}{$s$}}
    \COval(315,45)(7,7)(0){Black}{Black}
    \Text(330,45)[lb]{\scalebox{0.8}{$s$}}
    \Text(345,65)[lb]{\scalebox{1}{$-$}}
    \Line[](426,130)(426,95)
    \Line[](426,95)(367,18)
    \Line[](426,95)(486,18)
    \Line[](368,18)(486,18)
    \CBox(365,15)(370,20){Black}{Black}
    \CBox(482,15)(487,20){Black}{Black}
    \COval(426,93)(7,7)(0){Black}{Black}
    \COval(426,113)(7,7)(0){Black}{Black}
    \COval(457,55)(7,7)(0){Black}{Black}
    \Vertex(426,130){2}
    \Text(440,113)[lb]{\scalebox{0.8}{$s$}}
    \Text(470,55)[lb]{\scalebox{0.8}{$s$}}
    \Text(380,55)[lb]{\scalebox{0.8}{$s$}}
    \Text(425,3)[lb]{\scalebox{0.8}{$s$}}
    \Text(411,131)[lb]{\scalebox{0.8}{$x_3$}}
    \Text(356,5)[lb]{\scalebox{0.8}{$x_1$}}
    \Text(491,5)[lb]{\scalebox{0.8}{$x_2$}}
    \COval(425,18)(7,7)(0){Black}{Black}
    \COval(395,55)(7,7)(0){Black}{Black}
  \end{picture}
\end{center}
Notice that the first diagram is completely integrable, giving
\begin{equation}
\langle s(x_1)^2s(x_2)^2s(x_3)\rangle \supset \frac{C_{s^2 s^2 s}^{(0)}(1+\delta C_{s^2ss}-2A_s + v_2)\mu^{-\gamma_{s^2}-3\gamma_s}}
    {|x_{12}|^{6+\gamma_{s^2}+\gamma_s}|x_{13}|^{2-\gamma_s+\gamma_{s^2}}
    |x_{23}|^{2-\gamma_{s^2}+3\gamma_s}}\,,
\end{equation}
where
\begin{align}
v_2 {=} \frac{1}{N}\,\frac{\Gamma (d) \left(\left(3 (d{-}4) (d{-}2)
H_{d{-}4}{+}d ((d{-}15) d{+}60){-}60\right) \sin
\left(\frac{\pi  d}{2}\right){+}3 \pi  (d{-}4) (d{-}2) \cos \left(\frac{\pi  d}{2}\right)\right)}
{\pi  (d{-}4) \Gamma \left(\frac{d}{2}{+}1\right) \Gamma \left(\frac{d}{2}\right)}\,.
\end{align}
is obtained by the expansion of the $U$ functions generated during the integration over the unique internal vertices
\begin{align}
U\left(\frac{d-\gamma_s}{2}-1,\frac{d-\gamma_s}{2}-1,2+\gamma_s\right)
&U\left(1+\frac{\gamma_s}{2},2+\gamma_s,d-3-\frac{3\gamma_s}{2}\right)\\
&\times U\left(\frac{d-\gamma_s}{2}-1,2+\frac{\gamma_{s^2}}{2},
\frac{d+\gamma_s-\gamma_{s^2}}{2}-1\right)\,,\notag
\end{align}
while the second diagram is given by (\ref{sss correction in s2s2s}), for the total of
\begin{equation}
\label{s2s2s 1}
\langle s(x_1)^2s(x_2)^2s(x_3)\rangle \supset \frac{C_{s^2 s^2 s}^{(0)}(1+\delta C_{s^2ss}-2A_s + v_2-R_{sss})\mu^{-\gamma_{s^2}+2\gamma_s}}
    {|x_{12}|^{6+\gamma_{s^2}-2\gamma_s}|x_{13}|^{2-2\gamma_s+\gamma_{s^2}}
    |x_{23}|^{2-\gamma_{s^2}+2\gamma_s}}
    -\frac{C_{s^2 s^2 s}^{(0)}}{|x_{12}|^6(|x_{13}||x_{23}|)^4}\,,
\end{equation}
Analogously, correcting the right-hand $s^2 ss$ sub-diagram in (\ref{s2ss corrections in s2s2s}) we obtain
\begin{equation}
\label{s2s2s 2}
\langle s(x_1)^2s(x_2)^2s(x_3)\rangle \supset \frac{C_{s^2 s^2 s}^{(0)}(1+\delta C_{s^2ss}-2A_s + v_3-R_{sss})\mu^{-\gamma_{s^2}+2\gamma_s}}
    {|x_{12}|^{6+\gamma_{s^2}-2\gamma_s}|x_{13}|^{2-\gamma_{s^2}+2\gamma_s}
    |x_{23}|^{2-2\gamma_s+\gamma_{s^2}}}
    -\frac{C_{s^2 s^2 s}^{(0)}}{|x_{12}|^6(|x_{13}||x_{23}|)^4}\,,
\end{equation}
where 
\begin{equation}
v_3 = v_2\,.
\end{equation}

Combining (\ref{total sss correction in s2s2s}), (\ref{s2s2s 1}), (\ref{s2s2s 2}), we obtain
\begin{equation}
\label{w0 structure of s2s2s}
\langle s(x_1)^2s(x_2)^2s(x_3)\rangle \supset\frac{C_{s^2 s^2 s}^{(0)}(1+w_0)\mu^{-\gamma_s-2\gamma_{s^2}}}
    {|x_{12}|^{6+2\gamma_{s^2}-\gamma_s}(|x_{13}||x_{23}|)^{2+\gamma_s}}\,,
\end{equation}
where we denoted
\begin{equation}
w_0 = v_1+ 2v_2+2\delta C_{s^2ss}-4A_s-2R_{sss}\,.
\end{equation}
Using (\ref{v1}) we obtain
\begin{equation}
\label{w0}
w_0 = \delta C_{sss}-3A_s+2(\delta C_{s^2ss}+v_2-R_{sss})\,.
\end{equation}
Notice that (\ref{w0 structure of s2s2s}) already has the structure required by the conformal symmetry of 
the $\langle s^2 s^2 s\rangle$ three-point function (\ref{s2s2s general}). This means that the rest of the
diagrams which contribute to the $\langle s^2 s^2 s\rangle$ are finite, as we will confirm explicitly below in this section.
Expanding around $d=3$, we obtain
\begin{align}
v_2 = \frac{16}{\pi^2}\frac{1}{d-3}+{\cal O}((d-3)^0)\,,\qquad
R_{sss}= \frac{16}{\pi^2}\frac{1}{d-3}+{\cal O}((d-3)^0)\,.
\end{align}
We also know that $\langle sss\rangle$ vanishes in $d=3$, and therefore
$\delta C_{sss} = {\cal O}((d-3)^0)$.
In addition, we know that
$A_s = {\cal O}((d-3)^0)$.
Finally, while we haven't calculated $\delta C_{s^2ss}$, we know that
$C_{s^2ss}^{(0)}\delta C_{s^2ss}$ should be finite in $3d$, and therefore,
since $C_{s^2ss}^{(0)}$ is finite, we conclude that $\delta C_{s^2ss}$ must be
finite in $3d$.\footnote{Actually we calculated $\delta C_{s^2ss}$ in $3d$ and showed
explicitly that it is finite.} Then from (\ref{w0}) we conclude that
\begin{equation}
w_0 = {\cal O}((d-3)^0)\,.
\end{equation}

Consider the following diagram
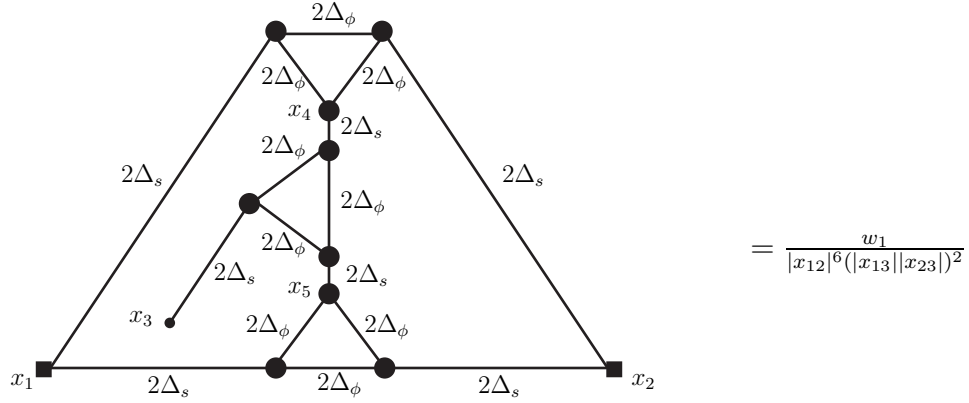
\begin{center}
  \begin{picture}(307,151) (23,11)
    \SetWidth{1.0}
    \SetColor{Black}
    \Line[](35,24)(245,24)
    \Line[](140,52)(119,24)
    \Line[](140,52)(161,24)
    \Line[](140,122)(119,150)
    \Line[](140,122)(161,150)
    \Line[](119,150)(161,150)
    \Line[](35,24)(119,150)
    \Line[](161,150)(245,24)
    \Line[](140,122)(140,52)
    \Line[](140,108)(112,87)
    \Line[](112,87)(140,66)
    \Line[](110,86)(80,41)
    \Vertex(140,52){4}
    \Vertex(120,24){4}
    \Vertex(161,24){4}
    \Vertex(140,66){4}
    \Vertex(110,86){4}
    \Vertex(140,106){4}
    \Vertex(140,121){4}
    \Vertex(120,151){4}
    \Vertex(160,151){4}
    \Vertex(80,41){2}
    \CBox(30,21)(35,26){Black}{Black}
    \CBox(245,21)(250,26){Black}{Black}
    \Text(62,93)[lb]{\scalebox{0.8}{$2\Delta_s$}}
    \Text(206,93)[lb]{\scalebox{0.8}{$2\Delta_s$}}
    \Text(197,13)[lb]{\scalebox{0.8}{$2\Delta_s$}}
    \Text(72,13)[lb]{\scalebox{0.8}{$2\Delta_s$}}
    \Text(136,13)[lb]{\scalebox{0.8}{$2\Delta_\phi$}}
    \Text(109,35)[lb]{\scalebox{0.8}{$2\Delta_\phi$}}
    \Text(154,35)[lb]{\scalebox{0.8}{$2\Delta_\phi$}}
    \Text(147,55)[lb]{\scalebox{0.8}{$2\Delta_s$}}
    \Text(97,56)[lb]{\scalebox{0.8}{$2\Delta_s$}}
    \Text(115,66)[lb]{\scalebox{0.8}{$2\Delta_\phi$}}
    \Text(115,103)[lb]{\scalebox{0.8}{$2\Delta_\phi$}}
    \Text(145,82)[lb]{\scalebox{0.8}{$2\Delta_\phi$}}
    \Text(145,111)[lb]{\scalebox{0.8}{$2\Delta_s$}}
    \Text(115,129)[lb]{\scalebox{0.8}{$2\Delta_\phi$}}
    \Text(153,129)[lb]{\scalebox{0.8}{$2\Delta_\phi$}}
    \Text(134,153)[lb]{\scalebox{0.8}{$2\Delta_\phi$}}
    \Text(20,16)[lb]{\scalebox{0.8}{$x_1$}}
    \Text(255,16)[lb]{\scalebox{0.8}{$x_2$}}
    \Text(65,40)[lb]{\scalebox{0.8}{$x_3$}}
    \Text(125,118)[lb]{\scalebox{0.8}{$x_4$}}
    \Text(125,51)[lb]{\scalebox{0.8}{$x_5$}}
    \Text(300,60)[lb]{\scalebox{1}{$=\frac{w_1}{|x_{12}|^6(|x_{13}||x_{23}|)^2}$}}
  \end{picture}
\end{center}
Integrating over all of the internal vertices except for $x_{4,5}$,
and denoting
\begin{equation}
\label{w1}
w_1 = 2^2\,N^2\left(-\frac{2}{\sqrt{N}}\right)^6 C_{sss}^{(0)}C_\phi^6C_s^4U\left(\frac{d}{2}-1,
\frac{d}{2}-1,2\right)^2U(1,2,d-3)^2\,\hat w_1\,,
\end{equation}
we obtain
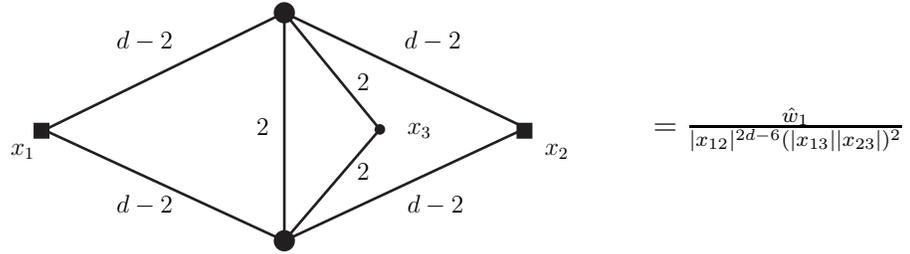
\begin{center}
  \begin{picture}(231,100) (36,-14)
    \SetWidth{1.0}
    \SetColor{Black}
    \Line[](43,33)(133,77)
    \Line[](133,-9)(42,33)
    \Line[](133,77)(223,34)
    \Line[](223,33)(133,-9)
    \CBox(39,30)(44,35){Black}{Black}
    \CBox(221,30)(226,35){Black}{Black}
    \Line[](133,77)(133,-9)
    \Line[](133,77)(169,33)
    \Line[](169,33)(133,-9)
    \Vertex(133,77){4}
    \Vertex(133,-9){4}
    \Vertex(169,33){2}
    \Text(30,22)[lb]{\scalebox{0.8}{$x_1$}}
    \Text(232,22)[lb]{\scalebox{0.8}{$x_2$}}
    \Text(180,30)[lb]{\scalebox{0.8}{$x_3$}}
    \Text(70,63)[lb]{\scalebox{0.8}{$d-2$}}
    \Text(179,63)[lb]{\scalebox{0.8}{$d-2$}}
    \Text(70,1)[lb]{\scalebox{0.8}{$d-2$}}
    \Text(180,1)[lb]{\scalebox{0.8}{$d-2$}}
    \Text(161,14)[lb]{\scalebox{0.8}{$2$}}
    \Text(161,48)[lb]{\scalebox{0.8}{$2$}}
    \Text(123,31)[lb]{\scalebox{0.8}{$2$}}
    \Text(273,25)[lb]{\scalebox{1}{$=\frac{\hat w_1}{|x_{12}|^{2d-6}(|x_{13}||x_{23}|)^2}$}}
  \end{picture}
\end{center}
Integrating both sides of the last diagrammatic equation w.r.t. $x_2$ we obtain
\begin{equation}
\label{hat w1}
\hat w_1 = \frac{\textrm{ChT}(1,1)}{U(1,2,d-3)}\,.
\end{equation}
Combining (\ref{w1}), (\ref{hat w1}), we obtain
\begin{equation}
\label{result for w1}
w_1{=}{-}\frac{1}{N^{3/2}}2^{4 d{-}3} (d{-}3)^6 \pi ^{{-}\frac{d}{2}{-}9}
\sin ^7\left(\frac{\pi  d}{2}\right) \Gamma \left(3{-}\frac{d}{2}\right)^3
\Gamma \left(\frac{d{-}3}{2}\right)^4 \left(\pi ^2{-}6 \psi ^{(1)}\left(\frac{d}{2}{-}1\right)\right)\,.
\end{equation}

Next, consider the diagram
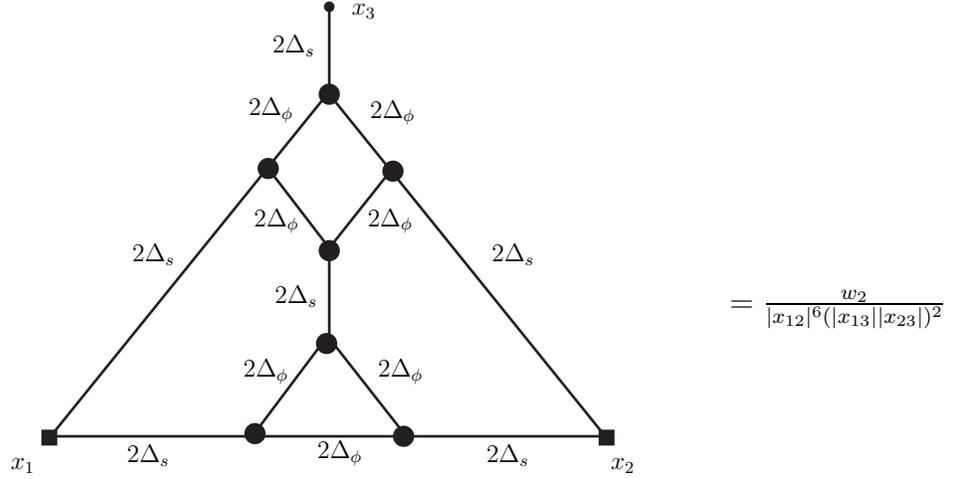
\begin{center}
  \begin{picture}(259,178) (31,10)
    \SetWidth{1.0}
    \SetColor{Black}
    \Line[](42,19)(252,19)
    \Line[](148,182)(148,148)
    \Line[](148,149)(43,19)
    \Line[](148,148)(252,19)
    \Line[](119,19)(146,55)
    \Line[](177,19)(148,56)
    \Line[](148,55)(148,90)
    \Line[](148,90)(171,119)
    \Line[](148,90)(125,120)
    \Text(28,5)[lb]{\scalebox{0.8}{$x_1$}}
    \Text(255,5)[lb]{\scalebox{0.8}{$x_2$}}
    \Text(157,177)[lb]{\scalebox{0.8}{$x_3$}}
    \Vertex(148,181){2}
    \Vertex(148,148){4}
    \Vertex(125,120){4}
    \Vertex(148,89){4}
    \Vertex(172,119){4}
    \Vertex(147,54){4}
    \Vertex(120,20){4}
    \Vertex(176,19){4}
    \CBox(40,16)(45,21){Black}{Black}
    \CBox(250,16)(255,21){Black}{Black}
    \Text(127,163)[lb]{\scalebox{0.8}{$2\Delta_s$}}
    \Text(118,138)[lb]{\scalebox{0.8}{$2\Delta_\phi$}}
    \Text(164,137)[lb]{\scalebox{0.8}{$2\Delta_\phi$}}
    \Text(120,96)[lb]{\scalebox{0.8}{$2\Delta_\phi$}}
    \Text(163,96)[lb]{\scalebox{0.8}{$2\Delta_\phi$}}
    \Text(74,84)[lb]{\scalebox{0.8}{$2\Delta_s$}}
    \Text(128,68)[lb]{\scalebox{0.8}{$2\Delta_s$}}
    \Text(210,84)[lb]{\scalebox{0.8}{$2\Delta_s$}}
    \Text(116,39)[lb]{\scalebox{0.8}{$2\Delta_\phi$}}
    \Text(167,39)[lb]{\scalebox{0.8}{$2\Delta_\phi$}}
    \Text(72,8)[lb]{\scalebox{0.8}{$2\Delta_s$}}
    \Text(144,8)[lb]{\scalebox{0.8}{$2\Delta_\phi$}}
    \Text(208,8)[lb]{\scalebox{0.8}{$2\Delta_s$}}
    \Text(300,60)[lb]{\scalebox{1}{$=\frac{w_2}{|x_{12}|^6(|x_{13}||x_{23}|)^2}$}}
  \end{picture}
\end{center}
It is straightforward to find $w_2$ by integrating both sides of this diagrammatic equation over $x_3$:
\begin{equation}
w_2 = 2^2N\left({-}\frac{2}{\sqrt{N}}\right)^4 C_{sss}^{(0)}C_\phi^4 C_s^3
U\left(\frac{d}{2}{-}1,\frac{d}{2}{-}1,2\right)^3U(2,2,d{-}4)\,.
\end{equation}
Simplifying this expression, we obtain
\begin{equation}
\label{result for w2}
w_2 =\frac{1}{N^{3/2}} \frac{4^{3 d-2} \sin ^6\left(\frac{\pi  d}{2}\right) \Gamma \left(2-\frac{d}{2}\right) \Gamma \left(4-\frac{d}{2}\right) \Gamma \left(\frac{d-1}{2}\right)^6}{\pi ^9 \Gamma (d-3)^2}\,.
\end{equation}

The other contributing diagram is given by
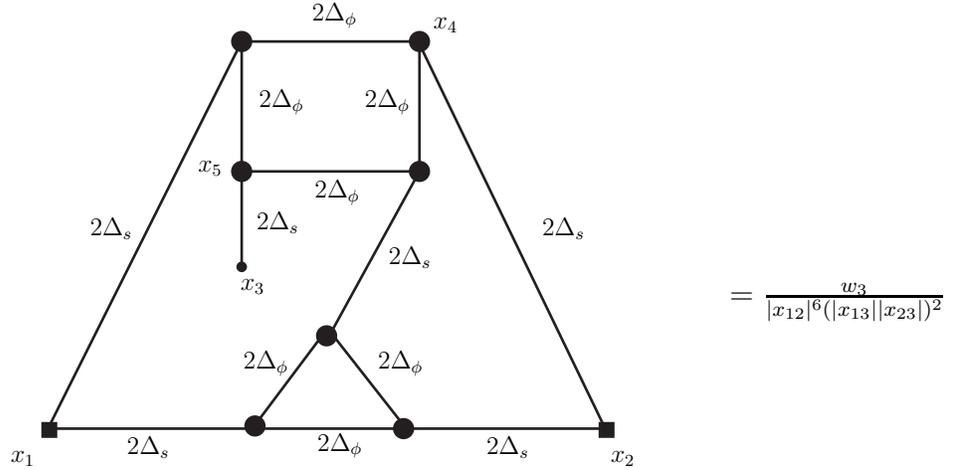
\begin{center}
  \begin{picture}(259,178) (31,10)
    \SetWidth{1.0}
    \SetColor{Black}
    \Line[](42,19)(252,19)
    \Line[](119,19)(146,55)
    \Line[](177,19)(148,56)
    \Text(28,5)[lb]{\scalebox{0.8}{$x_1$}}
    \Text(255,5)[lb]{\scalebox{0.8}{$x_2$}}
    \Vertex(147,54){4}
    \Vertex(120,20){4}
    \Vertex(176,19){4}
    \CBox(40,16)(45,21){Black}{Black}
    \CBox(250,16)(255,21){Black}{Black}
    \Text(116,39)[lb]{\scalebox{0.8}{$2\Delta_\phi$}}
    \Text(167,39)[lb]{\scalebox{0.8}{$2\Delta_\phi$}}
    \Text(72,8)[lb]{\scalebox{0.8}{$2\Delta_s$}}
    \Text(144,8)[lb]{\scalebox{0.8}{$2\Delta_\phi$}}
    \Text(208,8)[lb]{\scalebox{0.8}{$2\Delta_s$}}
    \Text(300,60)[lb]{\scalebox{1}{$=\frac{w_3}{|x_{12}|^6(|x_{13}||x_{23}|)^2}$}}
    \Line[](42,19)(115,165)
    \Line[](252,19)(182,164)
    \Line[](115,165)(182,165)
    \Line[](115,165)(115,116)
    \Line[](115,116)(182,116)
    \Line[](182,116)(182,165)
    \Line[](148,54)(182,116)
    \Line[](115,116)(115,80)
    \Vertex(115,165){4}
    \Vertex(182,165){4}
    \Vertex(115,116){4}
    \Vertex(182,116){4}
    \Vertex(115,80){2}
    \Text(142,171)[lb]{\scalebox{0.8}{$2\Delta_\phi$}}
    \Text(122,138)[lb]{\scalebox{0.8}{$2\Delta_\phi$}}
    \Text(162,138)[lb]{\scalebox{0.8}{$2\Delta_\phi$}}
    \Text(58,91)[lb]{\scalebox{0.8}{$2\Delta_s$}}
    \Text(229,91)[lb]{\scalebox{0.8}{$2\Delta_s$}}
    \Text(143,104)[lb]{\scalebox{0.8}{$2\Delta_\phi$}}
    \Text(121,93)[lb]{\scalebox{0.8}{$2\Delta_s$}}
    \Text(171,80)[lb]{\scalebox{0.8}{$2\Delta_s$}}
    \Text(115,70)[lb]{\scalebox{0.8}{$x_3$}}
    \Text(188,169)[lb]{\scalebox{0.8}{$x_4$}}
    \Text(99,115)[lb]{\scalebox{0.8}{$x_5$}}
  \end{picture}
\end{center}
To calculate this diagram, we proceed by integrating over the $sss$
sub-diagram, followed by the integral over the $x_{4,5}$ vertices.
Integrating both sides of the resulting diagrammatic equation over $x_1$, we
obtain
\begin{equation}
\label{w3}
w_3=2^2N\left({-}\frac{2}{\sqrt{N}}\right)^4 C_{sss}^{(0)}C_\phi^4 C_s^3
U\left(\frac{d}{2}{-}1,\frac{d}{2}{-}1,2\right)^2\,\hat w_3\,,
\end{equation}
where $\hat w_3$ is determined by
\begin{center}
  \begin{picture}(223,85) (41,-19)
    \SetWidth{1.0}
    \SetColor{Black}
    \Line[](43,20)(120,56)
    \Line[](120,56)(197,20)
    \Line[](197,20)(120,-14)
    \Line[](43,20)(120,-14)
    \Line[](120,56)(120,-14)
    \Vertex(120,56){4}
    \Vertex(120,-14){4}
    \CBox(195,17)(200,22){Black}{Black}
    \Vertex(43,20){2}
    \Text(27,20)[lb]{\scalebox{0.8}{$x_2$}}
    \Text(210,20)[lb]{\scalebox{0.8}{$x_3$}}
    \Text(70,45)[lb]{\scalebox{0.8}{$2$}}
    \Text(162,43)[lb]{\scalebox{0.8}{$2$}}
    \Text(128,20)[lb]{\scalebox{0.8}{$d-2$}}
    \Text(71,-7)[lb]{\scalebox{0.8}{$2$}}
    \Text(161,-6)[lb]{\scalebox{0.8}{$4$}}
    \Text(249,10)[lb]{\scalebox{1}{$=\frac{\hat w_3}{|x_{23}|^{8-d}}$}}
  \end{picture}
\end{center}
To calculate this self-energy diagram, we split the diagonal propagator into two merging propagators,
with the exponents $2d-4$ and $2$, while dividing the diagram by $U\left(\frac{d}{2}{-}1,\frac{d}{2}{-}1,2\right)$.
Integrating over the unique topmost vertex produces the factor of
$U\left(\frac{d}{2}{-}1,\frac{d}{2}{-}1,2\right)$, resulting in 
\begin{center}
  \begin{picture}(223,85) (41,-19)
    \SetWidth{1.0}
    \SetColor{Black}
    \Line[](43,20)(120,56)
    \Line[](120,56)(197,20)
    \Line[](197,20)(120,-14)
    \Line[](43,20)(120,-14)
    \Line[](120,56)(120,-14)
    \Vertex(120,56){4}
    \Vertex(120,-14){4}
    \CBox(195,17)(200,22){Black}{Black}
    \Vertex(43,20){2}
    \Text(27,20)[lb]{\scalebox{0.8}{$x_2$}}
    \Text(210,20)[lb]{\scalebox{0.8}{$x_3$}}
    \Text(55,45)[lb]{\scalebox{0.8}{$d-2+\delta$}}
    \Text(162,43)[lb]{\scalebox{0.8}{$d-2+\delta$}}
    \Text(128,20)[lb]{\scalebox{0.8}{$2$}}
    \Text(71,-7)[lb]{\scalebox{0.8}{$2$}}
    \Text(161,-6)[lb]{\scalebox{0.8}{$4$}}
    \Text(249,10)[lb]{\scalebox{1}{$=\frac{\hat w_3(\delta)}{|x_{23}|^{4+2\delta}}$}}
  \end{picture}
\end{center}
Here we introduced an auxiliary regulator $\delta$, which we will eventually set to zero.\footnote{The regulator
is auxiliary because the diagram is in fact finite in the $\delta\rightarrow 0$ limit.} Applying the integration by
parts identity to the topmost vertex (with the exponents
$\alpha_1 = 1$, $\alpha_2 = \alpha_3 = \frac{d+\delta}{2}-1$), we can solve for $\hat w_3(\delta)$ by calculating
four diagrams, three of which can be easily calculated using the propagator merging relations,
while the fourth one is the Fourier dual of the ChT diagram. Assembling everything 
together, and taking the limit $\delta\rightarrow 0$, we obtain
\begin{equation}
\hat w_3 = \frac{\pi ^{d+1}\left(-\pi ^2 (d-4)^2+\frac{8 (d ((d-4) d-8)+36)}
{(d-2)^2}+6 (d-4)^2 \psi ^{(1)}\left(\frac{d}{2}\right)\right)}{4 (d-4) \Gamma (d-2) \sin \left(\frac{\pi  d}{2}\right) }\,.
\end{equation}
Plugging this into (\ref{w3}) we obtain
\begin{align}
w_3&=\frac{1}{N^{3/2}}
\frac{2^{5 d-7} \sin ^5\left(\frac{\pi  d}{2}\right) \Gamma \left(3-\frac{d}{2}\right) \Gamma \left(\frac{d-1}{2}\right)^5 
}{\pi ^{15/2} \Gamma (d-3) \Gamma \left(\frac{d}{2}\right)^2}
\left(\pi ^2 \left(d^2-6 d+8\right)^2-6 \left(d^2-6 d+8\right)^2 \psi ^{(1)}\left(\frac{d}{2}\right)\right.\notag\\
&-\left. 8 (d ((d-4) d-8)+36)\right)\,.
\label{result for w3}
\end{align}

Next, consider the first non-planar pentagon-based diagram 
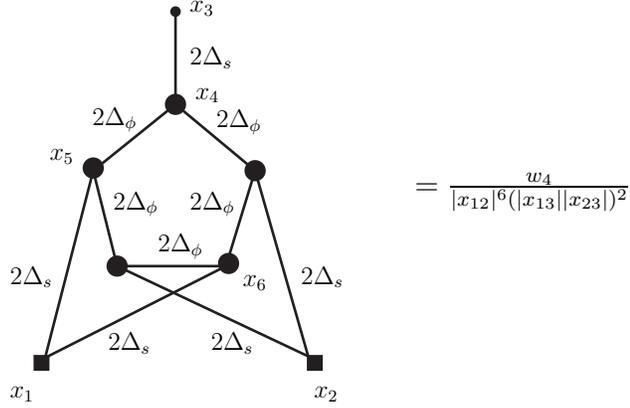
\begin{center}
  \begin{picture}(187,143) (73,5)
    \SetWidth{1.0}
    \SetColor{Black}
    \Line[](134,145)(134,110)
    \Line[](112,49)(153,49)
    \Line[](103,85)(112,49)
    \Line[](164,85)(153,49)
    \Line[](103,85)(134,110)
    \Line[](134,110)(164,85)
    \Line[](103,85)(85,15)
    \Line[](86,15)(153,49)
    \Line[](164,85)(184,15)
    \Line[](184,15)(112,49)
    \Vertex(103,86){4}
    \Vertex(112,49){4}
    \Vertex(154,50){4}
    \Vertex(164,85){4}
    \Vertex(134,110){4}
    \Vertex(134,145){2}
    \CBox(81,10)(86,15){Black}{Black}
    \CBox(184,10)(189,15){Black}{Black}
    \Text(103,100)[lb]{\scalebox{0.8}{$2\Delta_\phi$}}
    \Text(150,99)[lb]{\scalebox{0.8}{$2\Delta_\phi$}}
    \Text(111,68)[lb]{\scalebox{0.8}{$2\Delta_\phi$}}
    \Text(140,68)[lb]{\scalebox{0.8}{$2\Delta_\phi$}}
    \Text(128,52)[lb]{\scalebox{0.8}{$2\Delta_\phi$}}
    \Text(72,41)[lb]{\scalebox{0.8}{$2\Delta_s$}}
    \Text(182,41)[lb]{\scalebox{0.8}{$2\Delta_s$}}
    \Text(109,16)[lb]{\scalebox{0.8}{$2\Delta_s$}}
    \Text(148,16)[lb]{\scalebox{0.8}{$2\Delta_s$}}
    \Text(140,124)[lb]{\scalebox{0.8}{$2\Delta_s$}}
    \Text(225,70)[lb]{\scalebox{1}{$=\frac{w_4}{|x_{12}|^6(|x_{13}||x_{23}|)^2}$}}
    \Text(140,144)[lb]{\scalebox{0.8}{$x_3$}}
    \Text(72,-2)[lb]{\scalebox{0.8}{$x_1$}}
    \Text(187,-2)[lb]{\scalebox{0.8}{$x_2$}}
    \Text(142,111)[lb]{\scalebox{0.8}{$x_4$}}
    \Text(87,88)[lb]{\scalebox{0.8}{$x_5$}}
    \Text(160,40)[lb]{\scalebox{0.8}{$x_6$}}
  \end{picture}
\end{center}
To calculate this diagram, we integrate over the internal vertex $x_{4}$,
and then integrate both sides of this diagrammatic equation over $x_3$,
followed by integration over the vertices $x_{5,6}$. This will give
\begin{equation}
\label{w4}
w_4=2^2N\left({-}\frac{2}{\sqrt{N}}\right)^5 C_\phi^5 C_s^5
U\left(\frac{d}{2}{-}1,\frac{d}{2}{-}1,2\right)^3\,\hat w_4\,,
\end{equation}
where $\hat w_4$ is determined by
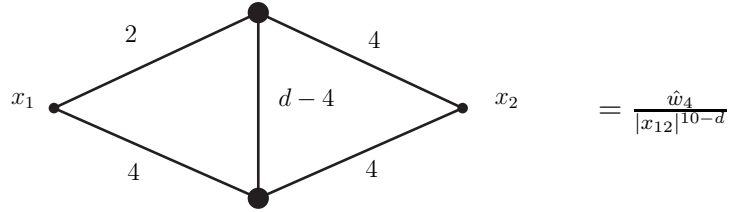
\begin{center}
  \begin{picture}(223,85) (41,-19)
    \SetWidth{1.0}
    \SetColor{Black}
    \Line[](43,20)(120,56)
    \Line[](120,56)(197,20)
    \Line[](197,20)(120,-14)
    \Line[](43,20)(120,-14)
    \Line[](120,56)(120,-14)
    \Vertex(120,56){4}
    \Vertex(120,-14){4}
    \Vertex(197,20){2}
    \Vertex(43,20){2}
    \Text(27,20)[lb]{\scalebox{0.8}{$x_1$}}
    \Text(210,20)[lb]{\scalebox{0.8}{$x_2$}}
    \Text(70,45)[lb]{\scalebox{0.8}{$2$}}
    \Text(162,43)[lb]{\scalebox{0.8}{$4$}}
    \Text(128,20)[lb]{\scalebox{0.8}{$d-4$}}
    \Text(71,-7)[lb]{\scalebox{0.8}{$4$}}
    \Text(161,-6)[lb]{\scalebox{0.8}{$4$}}
    \Text(249,10)[lb]{\scalebox{1}{$=\frac{\hat w_4}{|x_{12}|^{10-d}}$}}
  \end{picture}
\end{center}
To find $\hat w_4$, we split the diagonal propagator with the exponent $d-4$ into two propagators
with the exponents $2d-6$ and $2$, while dividing the diagram by $U(1,2,d-3)$. This will make
the topmost vertex unique, integrating over which will produce the factor of
$U(1,2,d-3)$, resulting in 
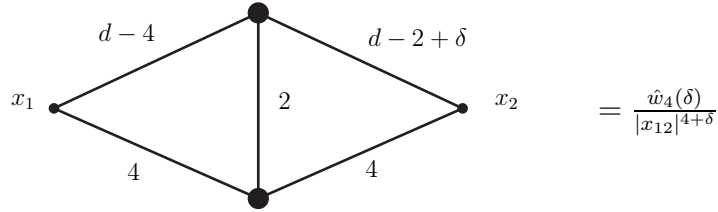
\begin{center}
  \begin{picture}(223,85) (41,-19)
    \SetWidth{1.0}
    \SetColor{Black}
    \Line[](43,20)(120,56)
    \Line[](120,56)(197,20)
    \Line[](197,20)(120,-14)
    \Line[](43,20)(120,-14)
    \Line[](120,56)(120,-14)
    \Vertex(120,56){4}
    \Vertex(120,-14){4}
    \Vertex(197,20){2}
    \Vertex(43,20){2}
    \Text(27,20)[lb]{\scalebox{0.8}{$x_1$}}
    \Text(210,20)[lb]{\scalebox{0.8}{$x_2$}}
    \Text(60,45)[lb]{\scalebox{0.8}{$d-4$}}
    \Text(162,43)[lb]{\scalebox{0.8}{$d-2+\delta$}}
    \Text(128,20)[lb]{\scalebox{0.8}{$2$}}
    \Text(71,-7)[lb]{\scalebox{0.8}{$4$}}
    \Text(161,-6)[lb]{\scalebox{0.8}{$4$}}
    \Text(249,10)[lb]{\scalebox{1}{$=\frac{\hat w_4(\delta)}{|x_{12}|^{4+\delta}}$}}
  \end{picture}
\end{center}
Here we introduced an auxiliary regulator $\delta$, which allows us to apply
the integration by parts relation to the topmost vertex (with the exponents $\alpha_1 = 1$,
$\alpha_2 = \frac{d}{2}-2$, $\alpha_3 = \frac{d+\delta}{2} - 1$), resulting in
\begin{align}
\label{hat w4}
\hat w_4 &= \frac{\frac{d}{2}-2}{1-\frac{\delta}{2}}\left(U(2,2,d-4)U\left(
\frac{d}{2}-1,\frac{d+\delta}{2}-1,2-\frac{\delta}{2}\right)-c_1\right)\\
&+\frac{\frac{d+\delta}{2}-1}{1-\frac{\delta}{2}}\left(U(2,2,d-4)U\left(
\frac{d}{2}-2,\frac{d+\delta}{2},2-\frac{\delta}{2}\right)-c_2\right)\,,\notag
\end{align}
where we denoted
\begin{center}
  \begin{picture}(223,85) (41,-19)
    \SetWidth{1.0}
    \SetColor{Black}
    \Line[](43,20)(120,56)
    \Line[](120,56)(197,20)
    \Line[](197,20)(120,-14)
    \Line[](43,20)(120,-14)
    \Line[](120,56)(120,-14)
    \Vertex(120,56){4}
    \Vertex(120,-14){4}
    \Vertex(197,20){2}
    \Vertex(43,20){2}
    \Text(27,20)[lb]{\scalebox{0.8}{$x_1$}}
    \Text(210,20)[lb]{\scalebox{0.8}{$x_2$}}
    \Text(65,45)[lb]{\scalebox{0.8}{$d-2$}}
    \Text(162,43)[lb]{\scalebox{0.8}{$d-2+\delta$}}
    \Text(128,20)[lb]{\scalebox{0.8}{$2$}}
    \Text(71,-7)[lb]{\scalebox{0.8}{$2$}}
    \Text(161,-6)[lb]{\scalebox{0.8}{$4$}}
    \Text(249,10)[lb]{\scalebox{1}{$=\frac{c_1}{|x_{12}|^{4+\delta}}$}}
  \end{picture}
\end{center}
\begin{center}
  \begin{picture}(223,85) (41,-19)
    \SetWidth{1.0}
    \SetColor{Black}
    \Line[](43,20)(120,56)
    \Line[](120,56)(197,20)
    \Line[](197,20)(120,-14)
    \Line[](43,20)(120,-14)
    \Line[](120,56)(120,-14)
    \Vertex(120,56){4}
    \Vertex(120,-14){4}
    \Vertex(197,20){2}
    \Vertex(43,20){2}
    \Text(27,20)[lb]{\scalebox{0.8}{$x_1$}}
    \Text(210,20)[lb]{\scalebox{0.8}{$x_2$}}
    \Text(65,45)[lb]{\scalebox{0.8}{$d-4$}}
    \Text(162,43)[lb]{\scalebox{0.8}{$d+\delta$}}
    \Text(128,20)[lb]{\scalebox{0.8}{$2$}}
    \Text(71,-7)[lb]{\scalebox{0.8}{$4$}}
    \Text(161,-6)[lb]{\scalebox{0.8}{$2$}}
    \Text(249,10)[lb]{\scalebox{1}{$=\frac{c_2}{|x_{12}|^{4+\delta}}$}}
  \end{picture}
\end{center}
The diagram $c_1$ is in fact finite, so we can set $\delta = 0$. In fact, it is equal to the
diagram $\hat w_3$, calculated above,
\begin{equation}
\label{c1}
c_1 = \lim_{\delta\rightarrow 0}\hat w_3\,.
\end{equation}
To calculate $c_2$, we apply the integration by parts relation to the topmost vertex
(with the exponents $\alpha_1 = 1$, $\alpha_2 = \frac{d}{2}-2$, $\alpha_3 = \frac{d+\delta}{2}$),
which will express that diagram in terms of a sum of four diagrams. Three of these diagrams can be easily
calculated using the propagator merging relation, while the fourth diagram is given by the Fourier
transform of the ChT diagram. Assembling everything together, we obtain
\begin{align}
c_2 &{=} \frac{4 (d{-}6) \pi ^{d+1} \csc \left(\frac{\pi  d}{2}\right)}{(d{-}2) \delta  \Gamma (d{-}3)}
{+}\frac{\pi^d}{12}\left({-}\frac{4 \left(\pi ^2 (d{-}4) (d{-}2){-}6 (d{-}4) d{+}24\right) \cos \left(\frac{\pi  d}{2}\right)
\Gamma (4{-}d)}{(d{-}2)^2}\right.\notag\\
&+\left.\frac{\pi  \csc \left(\frac{\pi  d}{2}\right) \Gamma (d-2)}{(d-4) \Gamma (d-1)^2}
\left(-18 \left(d^2-6 d+8\right)^2 \psi ^{(1)}\left(\frac{d}{2}\right)+\pi ^2 (d-2) d (d-4)^2\right.\right.\notag\\
&-\left.\left. 12 (d (d ((d-13) d+64)-140)+120)
\right)\right)\,.
\label{c2}
\end{align}
Combining (\ref{w4}), (\ref{hat w4}), (\ref{c1}), (\ref{c2}), we obtain
\begin{align}
w_4 &{=} {-}\frac{1}{N^{3/2}}\frac{32^{d-1} \Gamma \left(\frac{d-1}{2}\right)^5\sin ^5\left(\frac{\pi  d}{2}\right)}
{\pi ^{15/2} \Gamma \left(\frac{d}{2}-2\right)^2 \Gamma \left(\frac{d-2}{2}\right)}\left(-\frac{8 \cos \left(\frac{\pi  d}{2}\right) \Gamma (7-d)}{d^2-9 d+20}\right.\notag\\
&-\left.\frac{\pi  (d-4) \csc \left(\frac{\pi  d}{2}\right) \left(-8 d-6 \psi ^{(1)}\left(\frac{d}{2}-2\right)+\pi ^2+48\right)}{\Gamma (d-2)}\right)\,.
\label{result for w4}
\end{align}

The other non-planar diagram based on a pentagon effective vertex for $s$ is given by
\begin{center}
  \begin{picture}(343,83) (45,-22)
    \SetWidth{1.0}
    \SetColor{Black}
    \Line[](113,-11)(153,-11)
    \Line[](103,25)(113,-11)
    \Line[](165,25)(153,-11)
    \Line[](103,25)(135,50)
    \Line[](135,50)(165,25)
    \Vertex(103,25){4}
    \Vertex(113,-11){4}
    \Vertex(153,-11){4}
    \Vertex(165,25){4}
    \Vertex(134,50){4}
    \Line[](135,50)(273,50)
    \Line[](273,50)(153,-11)
    \Line[](165,25)(209,35)
    \Vertex(209,35){2}
    \Line[](103,25)(59,-11)
    \Line[](59,-11)(113,-11)
    \CBox(53,-14)(58,-9){Black}{Black}
    \CBox(272,47)(277,52){Black}{Black}
    \Text(196,54)[lb]{\scalebox{0.8}{$2\Delta_s$}}
    \Text(106,40)[lb]{\scalebox{0.8}{$2\Delta_\phi$}}
    \Text(152,37)[lb]{\scalebox{0.8}{$2\Delta_\phi$}}
    \Text(178,35)[lb]{\scalebox{0.8}{$2\Delta_s$}}
    \Text(66,10)[lb]{\scalebox{0.8}{$2\Delta_s$}}
    \Text(112,5)[lb]{\scalebox{0.8}{$2\Delta_\phi$}}
    \Text(140,5)[lb]{\scalebox{0.8}{$2\Delta_\phi$}}
    \Text(216,10)[lb]{\scalebox{0.8}{$2\Delta_s$}}
    \Text(81,-21)[lb]{\scalebox{0.8}{$2\Delta_s$}}
    \Text(130,-22)[lb]{\scalebox{0.8}{$2\Delta_\phi$}}
    \Text(42,-23)[lb]{\scalebox{0.8}{$x_1$}}
    \Text(283,47)[lb]{\scalebox{0.8}{$x_2$}}
    \Text(214,36)[lb]{\scalebox{0.8}{$x_3$}}
    \Text(171,17)[lb]{\scalebox{0.8}{$x_4$}}
    \Text(95,33)[lb]{\scalebox{0.8}{$x_5$}}
    \Text(111,-23)[lb]{\scalebox{0.8}{$x_6$}}
    \Text(303,20)[lb]{\scalebox{1}{$=\frac{w_5}{|x_{12}|^4(|x_{13}||x_{23}|)^2}$}}
  \end{picture}
\end{center}
To calculate this diagram, we integrate over $x_4$, followed by integration
of both sides of this diagrammatic equation over $x_3$, followed by integration
over $x_5$, $x_6$. This will give
\begin{equation}
\label{w5}
w_5=2^2N\left({-}\frac{2}{\sqrt{N}}\right)^5 C_\phi^5 C_s^5
U\left(\frac{d}{2}{-}1,\frac{d}{2}{-}1,2\right)^2
U\left(\frac{d}{2}{-}1,\frac{d}{2}{-}2,3\right)\,\hat w_5\,,
\end{equation}
where $\hat w_5$ is determined by
\begin{center}
  \begin{picture}(223,85) (41,-19)
    \SetWidth{1.0}
    \SetColor{Black}
    \Line[](43,20)(120,56)
    \Line[](120,56)(197,20)
    \Line[](197,20)(120,-14)
    \Line[](43,20)(120,-14)
    \Line[](120,56)(120,-14)
    \Vertex(120,56){4}
    \Vertex(120,-14){4}
    \Vertex(197,20){2}
    \Vertex(43,20){2}
    \Text(27,20)[lb]{\scalebox{0.8}{$x_1$}}
    \Text(210,20)[lb]{\scalebox{0.8}{$x_2$}}
    \Text(70,45)[lb]{\scalebox{0.8}{$4$}}
    \Text(162,43)[lb]{\scalebox{0.8}{$4$}}
    \Text(128,20)[lb]{\scalebox{0.8}{$d-6$}}
    \Text(71,-7)[lb]{\scalebox{0.8}{$4$}}
    \Text(161,-6)[lb]{\scalebox{0.8}{$4$}}
    \Text(249,10)[lb]{\scalebox{1}{$=\frac{\hat w_5}{|x_{12}|^{10-d}}$}}
  \end{picture}
\end{center}
To determine $\hat w_5$, we split the diagonal propagator with the
exponent $d-6$ into two, with the exponents $2d-8$ and $2$, while
dividing the diagram by $U(1,3,d-4)$. This will make the top-most vertex
unique, integrating over which will produce the factor of $U(2,2,d-4)$,
resulting in 
\begin{center}
  \begin{picture}(223,85) (41,-19)
    \SetWidth{1.0}
    \SetColor{Black}
    \Line[](43,20)(120,56)
    \Line[](120,56)(197,20)
    \Line[](197,20)(120,-14)
    \Line[](43,20)(120,-14)
    \Line[](120,56)(120,-14)
    \Vertex(120,56){4}
    \Vertex(120,-14){4}
    \Vertex(197,20){2}
    \Vertex(43,20){2}
    \Text(27,20)[lb]{\scalebox{0.8}{$x_1$}}
    \Text(210,20)[lb]{\scalebox{0.8}{$x_2$}}
    \Text(60,45)[lb]{\scalebox{0.8}{$d-4$}}
    \Text(162,43)[lb]{\scalebox{0.8}{$d-4$}}
    \Text(128,20)[lb]{\scalebox{0.8}{$2$}}
    \Text(71,-7)[lb]{\scalebox{0.8}{$4$}}
    \Text(161,-6)[lb]{\scalebox{0.8}{$4$}}
    \Text(249,10)[lb]{\scalebox{1}{$=\frac{\tilde w_5}{|x_{12}|^{2}}$}}
  \end{picture}
\end{center}
Here we denoted
\begin{equation}
\label{hat w5}
\hat w_5 = \frac{U(2,2,d-4)}{U(1,3,d-4)}\,\tilde w_5\,.
\end{equation}
To find $\tilde w_5$ we will apply the integration by parts
relation to the topmost vertex (with the exponents $\alpha_1 = 1$,
$\alpha_2 = \alpha_3 = \frac{d}{2}-2$), resulting in
\begin{equation}
\label{tilde w5}
\tilde w_5 = \frac{d-4}{4}\left(2U(2,2,d-4)U\left(\frac{d}{2}-1,\frac{d}{2}-2,3\right)-2c_3\right)\,,
\end{equation}
where $c_3$ is determined by
\begin{center}
  \begin{picture}(223,85) (41,-19)
    \SetWidth{1.0}
    \SetColor{Black}
    \Line[](43,20)(120,56)
    \Line[](120,56)(197,20)
    \Line[](197,20)(120,-14)
    \Line[](43,20)(120,-14)
    \Line[](120,56)(120,-14)
    \Vertex(120,56){4}
    \Vertex(120,-14){4}
    \Vertex(197,20){2}
    \Vertex(43,20){2}
    \Text(27,20)[lb]{\scalebox{0.8}{$x_1$}}
    \Text(210,20)[lb]{\scalebox{0.8}{$x_2$}}
    \Text(60,45)[lb]{\scalebox{0.8}{$d-4$}}
    \Text(162,43)[lb]{\scalebox{0.8}{$d-2+\delta$}}
    \Text(128,20)[lb]{\scalebox{0.8}{$2$}}
    \Text(71,-7)[lb]{\scalebox{0.8}{$4$}}
    \Text(161,-6)[lb]{\scalebox{0.8}{$2$}}
    \Text(249,10)[lb]{\scalebox{1}{$=\frac{c_3(\delta)}{|x_{12}|^{2+\delta}}$}}
  \end{picture}
\end{center}
Here we introduced an auxiliary regulator $\delta$, assuming that in
the end we will take the limit $c_3 =\lim_{ \delta\rightarrow 0} c_3(\delta)$.
To find $c_3(\delta)$ we apply the integration by parts relation to the top-most
vertex (with the exponents $\alpha_1 = 1$,
$\alpha_2 = \frac{d+\delta}{2}-1$, $\alpha_3 = \frac{d}{2}-2$), which will express that diagram in terms
of a sum of four diagrams. Three of these diagrams can be straightforwardly
integrated using the propagator merging relation, while the fourth one
is given by the Fourier transform of the ChT diagram. Assembling everything together,
we obtain
\begin{align}
\label{c3}
c_3 {=} -\frac{\pi ^{d+1} \csc \left(\frac{\pi  d}{2}\right) \left(\pi ^2 (d-4)^2-8 d-6 (d-4)^2 \psi ^{(1)}\left(\frac{d}{2}-1\right)+24\right)}{4 ((d-4) \Gamma (d-2))}\,.
\end{align}
Putting together (\ref{w5}), (\ref{hat w5}), (\ref{tilde w5}), (\ref{c3}), we obtain
\begin{align}
w_5 {=}\frac{1}{N^{3/2}}
 \frac{2^{4 d{-}1} \sin ^4\left(\frac{\pi  d}{2}\right) \Gamma \left(\frac{d{-}1}{2}\right)^4
 \left({-}\pi ^2 (d{-}4)^2{+}4 (d{-}3) (d{-}2){+}6 (d{-}4)^2 \psi ^{(1)}
 \left(\frac{d}{2}{-}1\right)\right)}{\pi ^6 (d{-}4)^2 \Gamma \left(\frac{d}{2}{-}2\right)^4}\,.
\label{result for w5}
\end{align}

The last diagram based on the pentagon effective vertex for $s$ is planar:
\begin{center}
  \begin{picture}(343,113) (45,-22)
    \SetWidth{1.0}
    \SetColor{Black}
    \Line[](113,-11)(153,-11)
    \Line[](103,25)(113,-11)
    \Line[](165,25)(153,-11)
    \Line[](103,25)(135,50)
    \Line[](135,50)(165,25)
    \Vertex(103,25){4}
    \Vertex(113,-11){4}
    \Vertex(153,-11){4}
    \Vertex(165,25){4}
    \Vertex(134,50){4}
    \Line[](134,50)(134,80)
    \Text(140,82)[lb]{\scalebox{0.8}{$x_3$}}
    \Text(140,63)[lb]{\scalebox{0.8}{$2\Delta_s$}}
    \Vertex(134,80){2}
    \Line[](103,25)(59,-11)
    \Line[](59,-11)(113,-11)
    \CBox(53,-14)(58,-9){Black}{Black}
    \Line[](207,-11)(153,-11)
    \CBox(203,-14)(208,-9){Black}{Black}
    \Line[](207,-11)(165,25)
    \Text(106,40)[lb]{\scalebox{0.8}{$2\Delta_\phi$}}
    \Text(152,37)[lb]{\scalebox{0.8}{$2\Delta_\phi$}}
    \Text(66,10)[lb]{\scalebox{0.8}{$2\Delta_s$}}
    \Text(112,5)[lb]{\scalebox{0.8}{$2\Delta_\phi$}}
    \Text(140,5)[lb]{\scalebox{0.8}{$2\Delta_\phi$}}
    \Text(80,-22)[lb]{\scalebox{0.8}{$2\Delta_s$}}
    \Text(130,-22)[lb]{\scalebox{0.8}{$2\Delta_\phi$}}
    \Text(175,-22)[lb]{\scalebox{0.8}{$2\Delta_s$}}
    \Text(186,10)[lb]{\scalebox{0.8}{$2\Delta_s$}}
    \Text(42,-23)[lb]{\scalebox{0.8}{$x_1$}}
    \Text(210,-23)[lb]{\scalebox{0.8}{$x_2$}}
    \Text(253,20)[lb]{\scalebox{1}{$=\frac{w_6}{|x_{12}|^4(|x_{13}||x_{23}|)^2}$}}
  \end{picture}
\end{center}
This diagram is straightforward to calculate by applying the uniqueness and the 
propagator merging relations, as well as integrating both sides over $x_3$, yielding
\begin{align}
w_6{=}2^2N\left({-}\frac{2}{\sqrt{N}}\right)^5 C_\phi^5 C_s^5
U\left(\frac{d}{2}{-}1,\frac{d}{2}{-}1,2\right)^3
U\left(\frac{d}{2}{-}1,\frac{d}{2}{-}2,3\right)U(2,3,d{-}5)\,.
\end{align}
Simplifying it, we obtain
\begin{equation}
\label{result for w6}
w_6 = -\frac{1}{N^{3/2}}\frac{2^{5 d-3} \sin ^5\left(\frac{\pi  d}{2}\right) \Gamma \left(5-\frac{d}{2}\right) \Gamma \left(\frac{d-1}{2}\right)^5}{\pi ^{15/2} (d-6)^3 \Gamma \left(\frac{d}{2}-1\right)^2 \Gamma (d-6)}\,.
\end{equation}

Combining everything together, we obtain
\begin{equation}
\boxed{
C_{s^2 s^2 s}^{(0)}\,\delta C_{s^2 s^2 s} = C_{s^2 s^2 s}^{(0)}\,w_0+ \sum_{a=1}^ 6 w_a
}
\end{equation}
The only non-vanishing contributions in $3d$ are given by
\begin{equation}
w_4(d=3) = \frac{1}{N^{3/2}}\,\frac{512}{\pi^6}\,,\qquad
w_5(d=3) = \frac{1}{N^{3/2}}\,\frac{256}{\pi^6}\,.
\end{equation}
This implies
\begin{equation}
C_{s^2 s^2 s}^{(0)}\,\delta C_{s^2 s^2 s}(d=3) =  \frac{1}{N^{3/2}}\,\frac{768}{\pi^6}\,.
\end{equation}

\section{Fate of the emergent $\mathbb{Z}_2$ symmetry at large $N$}
\label{sec:leading order}

In the previous section, we demonstrated that
while the leading ${\cal O}(1/N^{1/2})$ order amplitude of the $\langle s^2 s^2 s\rangle$
three-point function vanishes in $d=3$ dimensions,
its next-to-leading  ${\cal O}(1/N^{3/2})$ order correction is non-vanishing. This implies
that the conjectured $s\rightarrow -s$ symmetry of correlation functions in the $s$ sector
of the theory is violated at the first sub-leading order in $1/N$.

Instead, in this section we intend to discuss whether the statement of the $s\rightarrow -s$
symmetry has a more general validity at the leading order in $1/N$.
Specifically, we are going to focus on the three-point correlation functions $\langle s^k s^m s^n\rangle$,
with $k+m+n=2l+3$, where $k$, $m$, $n$ are positive integers and $l=0,1,2\dots$.
This question is particularly relevant from the standpoint of the holographic correspondence,
which is usually limited for technical reasons to the leading $1/N$ order calculations in the bulk \cite{Klebanov:2002ja,Petkou:2003zz,Sezgin:2003pt,Giombi:2009wh}.

Consider a sub-set of the $\langle s^k s^m s^n\rangle$ three-point functions that
at the leading order in $1/N$ expansion
are determined by the diagram
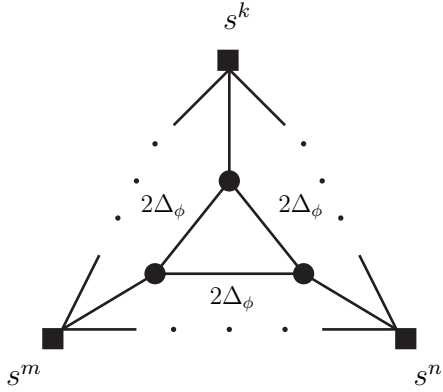
\begin{center}
  \begin{picture}(186,156) (38,14)
    \SetWidth{1.0}
    \SetColor{Black}
    \Line[](119,86)(91,51)
    \Line[](91,51)(147,51)
    \Line[](147,51)(119,86)
    \Line[](91,51)(56,30)
    \Line[](147,51)(182,30)
    \Line[](119,86)(119,128)
    \CBox(49,23)(56,30){Black}{Black}
    \CBox(182,23)(189,30){Black}{Black}
    \CBox(115,128)(122,135){Black}{Black}
    \Vertex(119,86){4}
    \Vertex(91,51){4}
    \Vertex(147,51){4}
    \Line[](119,128)(98,107)
    \Line[](119,128)(140,107)
    \Line[](56,30)(70,58)
    \Line[](56,30)(84,30)
    \Line[](182,30)(154,30)
    \Line[](182,30)(168,58)
    \Text(117,144)[lb]{\scalebox{1}{$s^k$}}
    \Text(35,9)[lb]{\scalebox{1}{$s^m$}}
    \Text(189,9)[lb]{\scalebox{1}{$s^n$}}
    \Text(86,72)[lb]{\scalebox{0.8}{$2\Delta_\phi$}}
    \Text(138,72)[lb]{\scalebox{0.8}{$2\Delta_\phi$}}
    \Text(112,37)[lb]{\scalebox{0.8}{$2\Delta_\phi$}}
    \Vertex(119,30){1}
    \Vertex(140,30){1}
    \Vertex(98,30){1}
    \Vertex(77,72){1}
    \Vertex(161,72){1}
    \Vertex(84,86){1}
    \Vertex(154,86){1}
    \Vertex(91,100){1}
    \Vertex(147,100){1}
  \end{picture}
\end{center}
Here we symbolically denoted the possible contractions of the $s$ lines with dots.
Since $k+m+n-3=2l$ is even-valued, we will always be able to connect the $s$ lines
which did not go into the vertices of the $\phi$ triangle.
To avoid tadpole loops of the $s$ propagators we need to impose conditions
\begin{align}
\boxed{
k\leq m+n-1\,,\quad k\geq m\geq n 
}
\label{triangle condition}
\end{align}
The resulting diagram will be given by\footnote{We skipped keeping track of numerical symmetry and degeneracy factors, that are unimportant for our purposes.}
\begin{equation}
\langle s^m(x_1) s^n(x_2) s^k(x_3)\rangle \simeq \frac{C_{sss}^{(0)}C_s^{l}}
{|x_{12}|^{2(m+n-k)}|x_{13}|^{2(m+k-n)}|x_{23}|^{2(k+n-m)}}\,.
\label{smsksn in 3d}
\end{equation}
Since $C_{sss}^{(0)}(d=3) = 0$, this leading ${\cal O}(1/N^{1/2})$
diagram (\ref{smsksn in 3d}) vanishes in $3d$. 

In case when the triangle inequalities (\ref{triangle condition}) are not satisfied,
the leading order behavior of the $\langle s^k s^m s^n\rangle$ three-point correlation functions
can be further suppressed than ${\cal O}(1/N^{1/2})$. For instance, the naive ${\cal O}(1/N^{1/2})$ diagram
contributing to $\langle s^3 s s\rangle$
is given by:\footnote{
To lighten up the notation, we skip labeling exponents of the propagators; all of the internal lines are the $s$
lines, except for the $\phi$ lines in the polygons.}
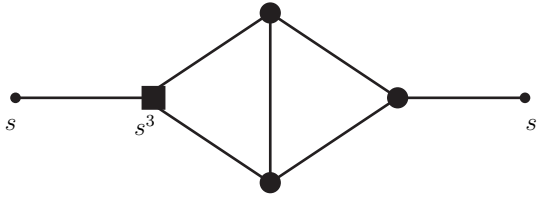
\begin{center}
  \begin{picture}(236,80) (27,5)
    \SetWidth{1.0}
    \SetColor{Black}
    \Line[](32,44)(80,44)
    \Line[](80,44)(128,76)
    \Line[](80,44)(128,12)
    \Line[](128,76)(128,12)
    \Line[](128,76)(176,44)
    \Line[](176,44)(128,12)
    \Line[](176,44)(224,44)
    \CBox(80,40)(88,48){Black}{Black}
    \Vertex(32,44){2}
    \Vertex(224,44){2}
    \Vertex(128,76){4}
    \Vertex(128,12){4}
    \Vertex(176,44){4}
    \Text(77,30)[lb]{\scalebox{0.8}{$s^3$}}
    \Text(225,32)[lb]{\scalebox{0.8}{$s$}}
    \Text(28,32)[lb]{\scalebox{0.8}{$s$}}
  \end{picture}
\end{center}
This  diagram is in fact identically zero in \textit{any} $d$.
While this can be established by an explicit calculation, a faster way to arrive
at the same conclusion is by noticing that the diagram above contains the $\langle s^2 s\rangle$
sub-diagram, which is zero due to conformal symmetry.
Along with the diagrams involving tadpoles of $s$, these kinds of diagrams do
not make any physical contributions to the corresponding correlation function.

As a result, the leading order contributions to the $\langle s^{3} s s\rangle$ 
will instead be of the order ${\cal O}(1/N^{3/2})$. Among the contributing diagrams
there are planar and non-planar graphs consisting of the pentagon of the $\phi$ lines, such as
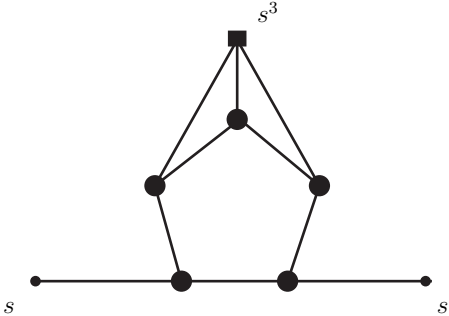
\begin{center}
  \begin{picture}(200,113) (45,-22)
    \SetWidth{1.0}
    \SetColor{Black}
    \Line[](113,-11)(153,-11)
    \Line[](103,25)(113,-11)
    \Line[](165,25)(153,-11)
    \Line[](103,25)(135,50)
    \Line[](135,50)(165,25)
    \Vertex(103,25){4}
    \Vertex(113,-11){4}
    \Vertex(153,-11){4}
    \Vertex(165,25){4}
    \Vertex(134,50){4}
    \Vertex(58,-11){2}
    \Vertex(205,-11){2}
    \Line[](134,50)(134,80)
    \CBox(131,78)(137,83){Black}{Black}
    \Line[](59,-11)(113,-11)
    \Line[](207,-11)(153,-11)
    \Line[](165,25)(134,80)
    \Line[](103,25)(134,80)
    \Text(142,87)[lb]{\scalebox{0.8}{$s^3$}}
    \Text(46,-23)[lb]{\scalebox{0.8}{$s$}}
    \Text(210,-23)[lb]{\scalebox{0.8}{$s$}}
  \end{picture}
\end{center}
Notice, that this diagram needs to be regularized.
It is an interesting open problem to finish the calculation of the
three-point function $\langle s^3 s s\rangle$, which is non-trivial
even at the leading ${\cal O}(1/N^{3/2})$ order.
Similar unsolved polygon diagrams appear in the leading order contributions
to the three-point functions $\langle s^ks^ms^n\rangle$ that do not 
satisfy the triangle inequalities  (\ref{triangle condition}). Therefore, an explicit proof of whether the $s\rightarrow -s$
symmetry persists throughout the entire set of these three-point correlators at the leading 
order in $1/N$ remains an open problem.

\section{Discussion}
\label{sec:discussion}

In this paper, we set out to calculate the three-point correlation function $\langle s^2s^2s\rangle$ in the critical $O(N)$ vector model at the next-to-leading order in the $1/N$ expansion. In the process, we computed the $s^2ss$ conformal triangle, following the technique developed in \cite{Goykhman:2020tsk} for the analogous calculation in the Gross-Neveu model. Additionally, we 
determined the finite correction $A_{s^2}$ to the amplitude of the $\langle s^2s^2\rangle$ two-point function. This involved calculating an extra diagram, absent in the GN model, that is hard to find in general $2<d<6$ using conventional techniques, and is still an open problem. However, it conveniently vanishes in $d=3$,\footnote{This can be traced back to the triviality of the $\langle sss \rangle$ sub-diagram, which appears in that calculation, at the leading order in $1/N$.} allowing us to compute the amplitude correction $A_{s^2}$ in $3d$. The computation of the $\langle s^2s^2s\rangle$ three-point function required evaluating a new set of self-energy diagrams that were unknown in the literature.
We have outlined the steps of all these calculations in detail. Assembling all these components together, we were able to compute the $\langle s^2s^2s\rangle$ correlator in $d=3$ dimensions, that turned out to have a non-zero value.

As was discussed in detail in section~\ref{sec:leading order}, we explored
 the $\mathbb{Z}_2$ symmetry $s\rightarrow -s$, that was originally conjectured to emerge in the large $N$ limit of the $O(N)$ vector model in $d=3$. In \cite{Goykhman:2019kcj}, the symmetry seemed to surprisingly persist in the three-point function $\langle sss\rangle$ up to the next-to-leading order in the $1/N$ expansion. The non-zero value at the next-to-leading order of the $\langle s^2s^2s\rangle$ correlator, obtained in this paper,
suggests that the  emergent $\mathbb{Z}_2$ symmetry is lifted at the sub-leading orders in $1/N$. Besides this result, the calculation of the conformal triangle $s^2ss$ as well as the sub-diagrams contributing to the $\langle s^2s^2s\rangle$ are
important in themselves for other conformal correlators in vector models.

Regarding the emergent symmetry, our results are supportive of the statement 
that it is exact at large $N$, as we have illustrated in section~\ref{sec:leading order}, where we demonstrated its presence for an entire set of correlators involving composite operators in $s$. This may have important implications for the AdS/CFT correspondence, which states that the critical $O(N)$ vector model in $3d$ is dual to a higher-spin Vasiliev theory on $AdS_4$. The AdS/CFT mapping suggests that the operator $s$ is dual to the spin zero field $A_0$ in AdS, and correspondingly all its polynomial powers correspond to the polynomial powers of $A_0$. Thus, any statement we make about large $N$ three-point correlators of $s^n$ operators in the boundary CFT, has a direct implication for the three-point correlators of $A_0^n$ in the bulk theory. Correlations of composite currents in the AdS bulk, are hitherto
largely undiscussed in the literature to the best of our knowledge.
However, symmetries of the bulk action alone, known to exist at least
at the classical level, can demand some boundary correlators to vanish. For instance,
the cubic interaction $A_0^3$ is absent in the Vasiliev's theory in the bulk \cite{Petkou:2003zz,Sezgin:2003pt},
which translates to the statement that the boundary CFT possesses the $s\rightarrow -s$
symmetry.
Next, the calculation of sub-leading corrections to such correlators in the large $N$ CFT has a direct implication on one-loop corrections in the AdS bulk, which are otherwise very hard to compute \cite{Aharony:2016dwx}. 
At the same time, the symmetry considerations might again indicate whether certain correlators can become non-zero
at sub-leading orders, due to an anomalous symmetry breaking mechanism in the bulk. An example
of the latter is furnished by the possible anomalous torsion term generated by loops in the bulk.\footnote{See, \textit{e.g.},  \cite{Leigh:2008tt} for a holographic description of parity-breaking system via torsion deformation of the AdS bulk. We thank A.~Petkou for the discussion and drawing our attention to relevant references.}

\section*{Acknowledgements} \noindent  We thank J.~Gracey, A.~Petkou, and especially M.~Smolkin for helpful discussions, comments on the draft,  and correspondence. Our work is partially supported by the Binational Science Foundation (grant No. 2016186), the Israeli Science Foundation Center of Excellence (grant No. 2289/18), and by the Quantum Universe I-CORE program of the Israel Planning and Budgeting Committee (grant No. 1937/12). The work of NC is partially supported by Yuri Milner scholarship.

\appendix

\appendix

\section{Some useful identities}
\label{appA}
In this appendix we review some useful expressions and identities.

Loop diagram in the position space are additive:
\begin{center}
  \begin{picture}(257,50) (0,0)
    \SetWidth{1.0}
    \SetColor{Black}
    \Arc[clock](81,-39)(77.006,127.614,52.386)
    \Arc[](80,86)(80,-126.87,-53.13)
    \Line[](160,22)(256,22)
    \Vertex(160,22){2}
    \Vertex(256,22){2}
    \Vertex(33,22){2}
    \Vertex(129,22){2}
    \Text(142,20)[lb]{$=$}
    \Text(77,-5)[lb]{\scalebox{0.8}{$2b$}}
    \Text(77,43)[lb]{\scalebox{0.8}{$2a$}}
    \Text(190,27)[lb]{\scalebox{0.8}{$2(a+b)$}}
  \end{picture}
\end{center}

The propagator merging relation has the form
\begin{equation}
\int d^d x_3\, \frac{1}{(x_3^2)^a((x_3-x_{12})^2)^b}
=U(a,b,d-a-b)\,\frac{1}{(x_{12}^2)^{a+b-\frac{d}{2}}}\,,
\end{equation}
where we defined
\begin{align}
U(a,b,c) &= \pi^\frac{d}{2} A(a)A(b)A(c)\,,\\
A(a)&=\frac{\Gamma\left(\frac{d}{2}-a\right)}{\Gamma(a)}\,.
\end{align}
This can be graphically represented as
\begin{center}
  \begin{picture}(98,10) (130,-60)
    \SetWidth{1.0}
    \SetColor{Black}
    \Line[](30,-58)(90,-58)
    \Line[](90,-58)(150,-58)
    \Line[](180,-58)(240,-58)
    \Vertex(30,-58){2}
    \Vertex(90,-58){4}
    \Vertex(150,-58){2}
    \Vertex(180,-58){2}
    \Vertex(240,-58){2}
    \Text(55,-53)[lb]{\scalebox{0.8}{$2a$}}
    \Text(115,-53)[lb]{\scalebox{0.8}{$2b$}}
    \Text(163,-61)[lb]{$=$}
    \Text(180,-53)[lb]{\scalebox{0.8}{$2(a+b)-d$}}
    \Text(250,-63)[lb]{$\times U(a,b,d-a-b)$}
  \end{picture}
\end{center}

Uniqueness relation for $a_1+a_2+a_3=d$ is written as \cite{DEramo:1971hnd,Symanzik:1972wj}
\begin{equation}
\int d^dx\,\frac{1}{|x_1-x|^{2a_1}|x_2-x|^{2a_2}|x_3-x|^{2a_3}}
=\frac{U(a_1,a_2,a_3)}{|x_{12}|^{d-2a_3}|x_{13}|^{d-2a_2}|x_{23}|^{d-2a_1}}\,,
\end{equation}
This can be diagrammatically represented as
\begin{center}
  \begin{picture}(210,66) (70,-31)
    \SetWidth{1.0}
    \SetColor{Black}
    \Line[](32,34)(80,2)
    \Line[](80,2)(32,-30)
    \Line[](80,2)(128,2)
    \Line[](192,34)(192,-30)
    \Line[](192,-30)(240,2)
    \Line[](240,2)(192,34)
    \Vertex(32,34){2}
    \Vertex(32,-30){2}
    \Vertex(80,2){4}
    \Vertex(128,2){2}
    \Vertex(192,34){2}
    \Vertex(192,-30){2}
    \Vertex(240,2){2}
    \Text(55,-25)[lb]{\scalebox{0.8}{$2a_1$}}
    \Text(55,22)[lb]{\scalebox{0.8}{$2a_2$}}
    \Text(100,5)[lb]{\scalebox{0.8}{$2a_3$}}
    \Text(160,-1)[lb]{$=$}
    \Text(180,-1)[lb]{\scalebox{0.8}{$\alpha$}}
    \Text(215,-28)[lb]{\scalebox{0.8}{$\beta$}}
    \Text(215,24)[lb]{\scalebox{0.8}{$\gamma$}}
    \Text(250,-5)[lb]{$\times \left(-\frac{2}{\sqrt{N}}\right) U\left(a_1,a_2,a_3\right)$}
  \end{picture}
\end{center}
Here we denoted $\alpha = d-2a_3$, $\beta = d-2a_2$, $\gamma = d-2a_1$.

Integration by parts relation \cite{Chetyrkin:1981qh,Kazakov:1983ns}:
\begin{center}
  \begin{picture}(250,192) (31,-130)
    \SetWidth{1.0}
    \SetColor{Black}
    \Line[](64,6)(32,-26)
    \Line[](64,6)(96,-26)
    \Line[](64,6)(64,54)
    \Vertex(64,6){4}
    \Text(50,27)[lb]{\scalebox{0.7}{$2\alpha_1$}}
    \Text(47,-20)[lb]{\scalebox{0.7}{$2\alpha_2$}}
    \Text(72,-20)[lb]{\scalebox{0.7}{$2\alpha_3$}}
    \Text(-40,3)[lb]{\scalebox{0.7}{$(d-2\alpha_1-\alpha_2-\alpha_3)\times$}}
    \Line[](174,6)(142,-26)
    \Line[](174,6)(206,-26)
    \Line[](174,6)(174,54)
    \Vertex(174,6){4}
    \Text(137,27)[lb]{\scalebox{0.7}{$2(\alpha_1-1)$}}
    \Text(114,-20)[lb]{\scalebox{0.7}{$2(\alpha_2+1)$}}
    \Text(182,-20)[lb]{\scalebox{0.7}{$2\alpha_3$}}
    \Text(120,3)[lb]{\scalebox{0.7}{$=\;\;\alpha_2\times\;\;$}}
    \Line[](284,6)(252,-26)
    \Line[](284,6)(316,-26)
    \Line[](284,6)(284,54)
    \Line[](252,-26)(284,54)
    \Vertex(284,6){4}
    \Text(290,27)[lb]{\scalebox{0.7}{$2\alpha_1$}}
    \Text(269,-20)[lb]{\scalebox{0.7}{$2(\alpha_2+1)$}}
    \Text(315,-20)[lb]{\scalebox{0.7}{$2\alpha_3$}}
    \Text(230,3)[lb]{\scalebox{0.7}{$-\;\;\alpha_2\times\;\;$}}
    \Text(255,20)[lb]{\scalebox{0.7}{$-2$}}
    \Line[](174,-94)(142,-126)
    \Line[](174,-94)(206,-126)
    \Line[](174,-94)(174,-46)
    \Vertex(174,-94){4}
    \Text(137,-73)[lb]{\scalebox{0.7}{$2(\alpha_1-1)$}}
    \Text(133,-120)[lb]{\scalebox{0.7}{$2\alpha_2$}}
    \Text(157,-120)[lb]{\scalebox{0.7}{$2(\alpha_3+1)$}}
    \Text(120,-97)[lb]{\scalebox{0.7}{$+\;\;\alpha_3\times\;\;$}}
    \Line[](284,-94)(252,-126)
    \Line[](284,-94)(316,-126)
    \Line[](284,-94)(284,-46)
    \Line[](316,-126)(284,-46)
    \Vertex(284,-94){4}
    \Text(267,-73)[lb]{\scalebox{0.7}{$2\alpha_1$}}
    \Text(243,-120)[lb]{\scalebox{0.7}{$2\alpha_2$}}
    \Text(267,-120)[lb]{\scalebox{0.7}{$2(\alpha_3+1)$}}
    \Text(230,-97)[lb]{\scalebox{0.7}{$-\;\;\alpha_3\times\;\;$}}
    \Text(300,-80)[lb]{\scalebox{0.7}{$-2$}}
  \end{picture}
\end{center}

We will also find useful the following relation \cite{Gracey:2018ame}\footnote{See fig. 15 therein.}
\begin{equation}
\label{selfenergy}
\scalebox{0.8}{
  \begin{picture}(223,285) (141,-219)
    \SetWidth{1}
    \SetColor{Black}
    \Line[](43,20)(120,56)
    \Line[](120,56)(197,20)
    \Line[](197,20)(120,-14)
    \Line[](43,20)(120,-14)
    \Line[](120,56)(120,-14)
    \Vertex(120,56){4}
    \Vertex(120,-14){4}
    \Vertex(197,20){2}
    \Vertex(43,20){2}
    \Text(27,20)[lb]{\scalebox{0.8}{$0$}}
    \Text(210,20)[lb]{\scalebox{0.8}{$x$}}
    \Text(60,45)[lb]{\scalebox{0.8}{$2\alpha_1$}}
    \Text(162,43)[lb]{\scalebox{0.8}{$2\alpha_2$}}
    \Text(128,20)[lb]{\scalebox{0.8}{$2\alpha_5$}}
    \Text(60,-7)[lb]{\scalebox{0.8}{$2\alpha_4$}}
    \Text(161,-6)[lb]{\scalebox{0.8}{$2\alpha_3$}}
    \Text(220,15)[lb]{\scalebox{1.2}{$=\frac{y_1}{|x|^2}\times$}}
    \Line[](293,20)(370,56)
    \Line[](370,56)(447,20)
    \Line[](447,20)(370,-14)
    \Line[](293,20)(370,-14)
    \Line[](370,56)(370,-14)
    \Vertex(370,56){4}
    \Vertex(370,-14){4}
    \Vertex(447,20){2}
    \Vertex(293,20){2}
    \Text(284,20)[lb]{\scalebox{0.8}{$0$}}
    \Text(460,20)[lb]{\scalebox{0.8}{$x$}}
    \Text(310,45)[lb]{\scalebox{0.8}{$2\alpha_1$}}
    \Text(412,43)[lb]{\scalebox{0.8}{$2\alpha_2$}}
    \Text(378,20)[lb]{\scalebox{0.8}{$2\alpha_5-2$}}
    \Text(310,-7)[lb]{\scalebox{0.8}{$2\alpha_4$}}
    \Text(411,-6)[lb]{\scalebox{0.8}{$2\alpha_3$}}
    \Text(220,-85)[lb]{\scalebox{1.2}{$+\frac{y_2}{|x|^2}\times$}}
    \Line[](293,-80)(370,-44)
    \Line[](370,-44)(447,-80)
    \Line[](447,-80)(370,-114)
    \Line[](293,-80)(370,-114)
    \Line[](370,-44)(370,-114)
    \Vertex(370,-44){4}
    \Vertex(370,-114){4}
    \Vertex(447,-80){2}
    \Vertex(293,-80){2}
    \Text(284,-80)[lb]{\scalebox{0.8}{$0$}}
    \Text(460,-80)[lb]{\scalebox{0.8}{$x$}}
    \Text(310,-55)[lb]{\scalebox{0.8}{$2\alpha_1$}}
    \Text(412,-57)[lb]{\scalebox{0.8}{$2\alpha_2$}}
    \Text(378,-80)[lb]{\scalebox{0.8}{$2\alpha_5$}}
    \Text(310,-107)[lb]{\scalebox{0.8}{$2\alpha_4$}}
    \Text(411,-106)[lb]{\scalebox{0.8}{$2\alpha_3-2$}}
    \Text(220,-185)[lb]{\scalebox{1.2}{$+\frac{y_3}{|x|^2}\times$}}
    \Line[](293,-180)(370,-144)
    \Line[](370,-144)(447,-180)
    \Line[](447,-180)(370,-214)
    \Line[](293,-180)(370,-214)
    \Line[](370,-144)(370,-214)
    \Vertex(370,-144){4}
    \Vertex(370,-214){4}
    \Vertex(447,-180){2}
    \Vertex(293,-180){2}
    \Text(284,-180)[lb]{\scalebox{0.8}{$0$}}
    \Text(460,-180)[lb]{\scalebox{0.8}{$x$}}
    \Text(310,-155)[lb]{\scalebox{0.8}{$2\alpha_1$}}
    \Text(412,-157)[lb]{\scalebox{0.8}{$2\alpha_2-2$}}
    \Text(378,-180)[lb]{\scalebox{0.8}{$2\alpha_5$}}
    \Text(310,-207)[lb]{\scalebox{0.8}{$2\alpha_4$}}
    \Text(411,-206)[lb]{\scalebox{0.8}{$2\alpha_3$}}
  \end{picture}
 }
\end{equation}
Here
\begin{align}
y_1&=\frac{(d-s_1)(d-s_2)}{(d-t_2)(t_2-d/2-1)}\,,\;\;
y_2=\frac{(d-s_2)(D+\alpha_5-3d/2-1)}{(d-t_2)(t_2-d/2-1)}\,,\\
y_3&=\frac{(d-s_1)(D+\alpha_5-3d/2-1)}{(d-t_2)(t_2-d/2-1)}\,,
\end{align}
and
\begin{align}
s_1 &=\alpha_1+\alpha_2+\alpha_5\,,\qquad
s_2 = \alpha_3+\alpha_4+\alpha_5\,,\\
t_1 &= \alpha_1+\alpha_4+\alpha_5\,,\qquad
t_2 =\alpha_2+\alpha_3+\alpha_5\,,\quad
D = \sum_{i=1}^5 \alpha_i\,.
\end{align}

\section{Some useful diagrams}

Using the integration by parts relation, we can derive  \cite{Chetyrkin:1981qh,Kazakov:1983ns,Goykhman:2019kcj}
\begin{center}
  \begin{picture}(223,85) (41,-19)
    \SetWidth{1.0}
    \SetColor{Black}
    \Line[](43,20)(120,56)
    \Line[](120,56)(197,20)
    \Line[](197,20)(120,-14)
    \Line[](43,20)(120,-14)
    \Line[](120,56)(120,-14)
    \Vertex(120,56){4}
    \Vertex(120,-14){4}
    \Vertex(197,20){2}
    \Vertex(43,20){2}
    \Text(27,20)[lb]{\scalebox{0.8}{$x_1$}}
    \Text(210,20)[lb]{\scalebox{0.8}{$x_2$}}
    \Text(75,45)[lb]{\scalebox{0.8}{$2$}}
    \Text(162,43)[lb]{\scalebox{0.8}{$2$}}
    \Text(128,20)[lb]{\scalebox{0.8}{$2$}}
    \Text(70,-7)[lb]{\scalebox{0.8}{$2\alpha$}}
    \Text(161,-6)[lb]{\scalebox{0.8}{$2\beta$}}
    \Text(249,10)[lb]{\scalebox{1}{$=\frac{F(\alpha,\beta)}{|x_{12}|^{6-2d+2\alpha+2\beta}}$}}
  \end{picture}
\end{center}
where
\begin{align}
F(\alpha,\beta) &{=}\frac{U(d{-}2,1,1)}{d{-}2{-}\alpha{-}\beta}\left(\alpha\left(
U(\alpha{+}1,\beta,d{-}\alpha{-}\beta{-}1){-}U\left(\alpha{+}1,\beta{+}2{-}\frac{d}{2},
\frac{3d}{2}{-}\alpha{-}\beta{-}3\right)\right)\right.\notag\\
&{+}\left.\beta\left(U(\beta{+}1,\alpha,d{-}\alpha{-}\beta{-}1){-}U\left(\beta{+}1,\alpha{+}2{-}\frac{d}{2},
\frac{3d}{2}{-}\alpha{-}\beta{-}3\right)\right)\right)\,.
\end{align}

Performing the Fourier transform we can derive the ChT diagram  \cite{Chetyrkin:1981qh,Kazakov:1983ns}
\begin{center}
  \begin{picture}(223,85) (41,-19)
    \SetWidth{1.0}
    \SetColor{Black}
    \Line[](43,20)(120,56)
    \Line[](120,56)(197,20)
    \Line[](197,20)(120,-14)
    \Line[](43,20)(120,-14)
    \Line[](120,56)(120,-14)
    \Vertex(120,56){4}
    \Vertex(120,-14){4}
    \Vertex(197,20){2}
    \Vertex(43,20){2}
    \Text(27,20)[lb]{\scalebox{0.8}{$x_1$}}
    \Text(210,20)[lb]{\scalebox{0.8}{$x_2$}}
    \Text(60,45)[lb]{\scalebox{0.8}{$d-2$}}
    \Text(162,43)[lb]{\scalebox{0.8}{$2\alpha$}}
    \Text(128,20)[lb]{\scalebox{0.8}{$d-2$}}
    \Text(60,-7)[lb]{\scalebox{0.8}{$d-2$}}
    \Text(161,-6)[lb]{\scalebox{0.8}{$2\beta$}}
    \Text(249,10)[lb]{\scalebox{1}{$=\frac{\textrm{ChT}(\alpha,\beta)}{|x_{12}|^{d-6+2\alpha+2\beta}}$}}
  \end{picture}
\end{center}
where
\begin{align}
\textrm{ChT}(\alpha,\beta) &= \frac{\pi ^d \Gamma \left(2-\frac{d}{2}\right)}{\Gamma \left(\frac{d}{2}-1\right) \Gamma (d-2)}\left(
\frac{\Gamma \left(\frac{d}{2}-\alpha \right) \Gamma \left(\frac{d}{2}+\alpha -2\right)}{(1-\beta ) (\alpha +\beta -2) \Gamma (2-\alpha ) \Gamma (\alpha )}\right.\\
&+\left.\frac{\Gamma \left(\frac{d}{2}-\beta \right) \Gamma \left(\frac{d}{2}+\beta -2\right)}{(1-\alpha ) (\alpha +\beta -2) \Gamma (2-\beta ) \Gamma (\beta )}+\frac{\Gamma \left(\frac{d}{2}-\alpha -\beta +1\right) \Gamma \left(\frac{d}{2}+\alpha +\beta -3\right)}{(\alpha -1) (\beta -1) \Gamma (-\alpha -\beta +3) \Gamma (\alpha +\beta -1)}\right)\,.\notag
\end{align}

\end{document}